\documentclass[%
reprint,
superscriptaddress,
showpacs,
amsmath,amssymb,amsfonts,
aps,
pra,
floatfix,
a4paper,
twocolumn
]{revtex4-2}
\usepackage{shellesc}% to compile with lualatex allowing latex to create files (necessary for tikzexternalize).
\usepackage{iftex}%to detect which compiler is being used and act as a switch
\ifluatex 	%detects if luatex is the compiler
	\usepackage{fontspec}	%instead of fontenc when using lualatex.
	\defaultfontfeatures{Ligatures=TeX}%to maintain tex ligatures even if changing the font.
\else
	\usepackage[utf8]{inputenc}%sets the encoding of the input text to the recent Unicode standard. (it includes accented characters and other alphabets, like cyrillic).
	\usepackage[T1]{fontenc}%sets the output font encoding to T1, which is one including accented characters. Some T2 encoding, like T2A, would need to be added to write in cyrillic. NOTE: the produced PDF may start using non-vectorial fonts. In order to solve this, install a font package including vectorial fonts for the new encoding into the latex distribution (e.g. MikTex), like cm-super. 
\fi
\DeclareMathOperator{\sgn}{sgn}%declares the sign operator
\DeclareMathOperator{\grad}{grad}%declares the gradient operator
\newcommand{\area}{\mathcal{A}}%declares the symbol for the generalized area
\newcommand{\energy}{\mathcal{E}}%declares the symbol for the pulse energy
\newcommand{\fidelity}{\mathcal{F}}%declares the symbol for the fidelity
\newcommand{\lagrangian}{\mathcal{L}}%declares the symbol for the lagrangian
\usepackage{graphicx}% Include figure files
\usepackage[usenames,dvipsnames]{xcolor}
\usepackage{dcolumn}% Align table columns on decimal point
\usepackage{bm}% bold math
\usepackage{bookmark}	%turns crossreferences, table of contents and other objects into hyperlinks. Also adds functionalities to configure the appearance of the document on the pdf viewer.
\usepackage[all]{hypcap}
\newcommand{\refsubcap}[2]{\hyperlink{#1}{\ref*{#1}#2}}
\makeatletter			%begins environment to execute native tex commands.
\newcommand*\nobreakhyphen{\hbox{-}\nobreak\hskip\z@skip}%allows hyphenation on words merged with hyphens.
\makeatother			%ends environment to execute native tex commands.
\usepackage{tikz}	    %Allows to make drawings directly with latex.
\usetikzlibrary{backgrounds}%let's you set background colors for the tikz-produced figures.
\usepackage{pgfplots}	%allows to make plots directly with latex.
\pgfkeys{/pgf/number format/.cd, set thousands separator={}}%suppresses the thousands separator.
\usepackage{pgfplotstable}%provides functionalities for the manipulation of tables.
\pgfplotsset{compat=newest}%sets the compatibility mode of pgfplots to the most recent version.
\usepgfplotslibrary{external}%makes latex to produce pdf versions of the figures made with tikz/pgfplots, which are then included in subsequent compilations in order to save time. %contains \usetikzlibrary{external} (and maybe adds something).
\usepgfplotslibrary{groupplots}%to use the environment groupplots to produce arrays of subplots.
\definecolor{mycolor1}{rgb}{0.00000,0.44700,0.74100}%blueish
\definecolor{mycolor2}{rgb}{0.85000,0.32500,0.09800}%redish
\definecolor{mycolor3}{rgb}{0.92900,0.69400,0.12500}%yellowish
\definecolor{mycolor4}{rgb}{0.49400,0.18400,0.55600}%purpleish
\definecolor{mycolor5}{rgb}{0.46600,0.67400,0.18800}%greenish
\definecolor{mycolor6}{rgb}{0.30100,0.74500,0.93300}%cyanish
\definecolor{mycolor7}{rgb}{0.63500,0.07800,0.18400}%maroonish
\definecolor{mybackground}{rgb}{1,1,1}%defines a variable with white background color.
\definecolor{myforeground}{rgb}{0,0,0}%defines a variable with a black foreground color (e.g. fonts)
\begin{document}
\title{Optimal robust stimulated Raman exact passage by inverse optimization}
\author{Xavier Laforgue}
\affiliation{Laboratoire Interdisciplinaire Carnot de Bourgogne, UMR CNRS 6303,  Universit\'e de Bourgogne Franche-Comt\'e, BP 47870, F-21078 Dijon, France}%
\author{Ghassen Dridi}
\affiliation{Institut Sup\'erieur des Sciences Appliqu\'ees et de Technologies de Gafsa, Universit\'e de Gafsa,
Campus Universitaire Sidi Ahmed Zarroug, Gafsa 2112, Tunisia}
\affiliation{Laboratoire de mat\'eriaux avanc\'es et ph\'enom\`enes quantiques. Universit\'e  de Tunis El Manar ll. Facult\'e des Sciences de Tunis, 2092, Tunisia}
\author{St\'ephane Gu\'erin}
\email{sguerin@u-bourgogne.fr}
\affiliation{Laboratoire Interdisciplinaire Carnot de Bourgogne, UMR CNRS 6303,  Universit\'e de Bourgogne Franche-Comt\'e, BP 47870, F-21078 Dijon, France}%
\date{\today}
\begin{abstract}
We apply the inverse geometric optimization technique to generate an optimal and robust stimulated Raman exact passage (STIREP) considering the loss of the upper state as a characterization parameter.
Control fields temporal shapes that are optimal with respect to pulse area, energy, and duration, are found to form a simple sequence with a combination of intuitively (near the beginning and the end) and counter-intuitively ordered pulse pairs.
%Robustness up to third order with respect to pulse amplitude inhomogeneities is demonstrated, accounting for a 13-fold increase on the fidelity profile waist at $\fidelity=10^{-4}$ regarding the unconstrained optimal.
%The resulting population dynamics produces a loss, through spontaneous emission, of $0.1291\Gamma T$ where $T$ is the interaction time and $\Gamma$ the dissipation rate; this represents a 34\% reduction from the non-robust optimal case. We find a family of optimal robust solutions featuring even lower losses that the optimal one, but with larger pulse areas.
The resulting dynamics produces a loss which is about a third of that of the non-robust optimal STIREP. Alternative optimal solutions featuring lower losses, larger pulse areas, and fully counter-intuitive pulse sequences are derived.
\end{abstract}
%\pacs{03.65.Aa, %Quantum systems with finite Hilbert space
%32.80.Qk, %Coherent control of atomic interactions with photons
%42.50.Dv %Quantum state engineering and measurements
%42.50.Ex %Optical implementations of quantum information processing and transfer}
\maketitle
%\listoftodos
\section{Introduction}
%It is well known that the populations of $N$-level quantum systems under constant (time-independent) fields, coupling resonantly some of said levels, undergo (Rabi) oscillations 
%\cite{Shore2011}.
Control of the population evolution of three-level lambda ($\Lambda$) systems has led to many applications, particularly in quantum control, due to its ubiquitous nature in the study of quantum processes \cite{Gaubatz1990,Kuhn2002,Sorensen2006,Kral2007,Bergmann2015,Vitanov2017}.
Such $\Lambda$ configurations are most relevant when the two quantum levels of interest, typically two stable ground states, are difficult, impossible, or impractical, to couple while there is a third, more energetic level, accessible from both others. 
We can transfer population between the two ground states, referred in short as $\Lambda$ transfer, using pump and Stokes controls, connected to the initial and target states, respectively, which produce Rabi oscillations when they fully overlap \cite[p.~197]{Shore2011}. 
However, this places significant transient population on the excited lossy state, which lead to an incomplete population transfer to the target state.
A way to overcome this difficulty was found in the technique known as stimulated Raman adiabatic passage (STIRAP) \cite{Bergmann2015}: a sequence of counter-intuitive pulses (Stokes pulse switched on before pump pulse with both pulses of the same duration) induces an adiabatic transfer with small transient population on the excited state when the pulse areas are large enough. 
Additionally, adiabaticity offers robustness, in particular, with respect to any specific design of the pulses. 

Adiabatic passage requires, in principle, infinite pulse areas to perform complete population transfers and to maintain the excited state completely depopulated along the dynamics, as it would be desirable.
Alternative protocols with realistic physical conditions have been recently investigated. 
One can mention acceleration of the transfer by parallel adiabatic passage \cite{Dridi2009}, but still featuring a large transient population in the excited state, shortcut to adiabaticity by counterdiabatic driving \cite{Ibanez2013,Song2016,Huang2016,Li2016,Kang2016}, and inverse engineering, where the controls are derived from a given dynamics \cite{Chen2012,Laforgue2019,Liu2019,Liu2021}. 
In the latter, the controls are determined from a prescribed dynamics chosen to achieve an exact transfer. 
This is referred to as stimulated Raman exact passage as opposed to the adiabatic, i.e.~approximate, passage \cite{Dorier2017}. 
The formulation allows one to generate infinitely many exact solutions and one can select among them the ones with specific features such as robustness \cite{Daems2013,Van-Damme2017,Laforgue2019,Zeng2018,Zeng2018a,Guengoerdue2019,Dong2021} or stability in the case of non-linear dynamics \cite{Dorier2017,Zhu2020}. 
Disadvantageously, this method is strongly dependent on the protocol used for the prescription, both in dynamical behavior and in the consequent field characteristics such as pulse area, energy, duration, and robustness.

A method combining inverse engineering and optimization, where the controls are derived from a trajectory that is optimal with respect to a given cost, has been proposed.
Robust inverse optimization (RIO) incorporates robustness as additional constraints.
It has been demonstrated for a two-level system using a variational procedure based on a geometric representation of the dynamics, producing ultimate solutions that featured exactness, robustness, and absolute optimality (for instance with respect to pulse area, energy, and duration of the controls) \cite{Dridi2020}.

%Parametrizing the propagator of the system, in the interaction representation, prescribing time-dependent functions for these parameters that will achieve the desired dynamics, and deduce the appropriate controls from these is known as reverse engineering and is a way to obtain complete population transfer.
%Transfers designed in this manner are then called exact, since they describe exactly the dynamics of the populations at any point in time and provide the exact final state that was targeted \cite{Laforgue2019}.

In this work, we apply the RIO technique to derive resonant control pulses in a $\Lambda$ system featuring exact, robust, and optimal transfers, taking into account a given admissible total loss. 
These result from an optimization via the resolution of the Euler-Lagrange equations with the constraints of robustness up to third order (considered in terms of a common scaling inhomogeneity factor for both fields). 
We find numerically the optimal and robust family of solutions, each of them corresponding to a given loss. 
These numerical solutions lead to control fields of remarkably simple temporal shapes featuring, when a low loss is considered, a combination of intuitively and counter-intuitively ordered pairs. 
Their area is only about twice as large as the optimal unconstrained (i.e.~non-robust) $\Lambda$ transfer \cite{Boscain2002,Boscain2002a}.

A simple guide to the results of this manuscript can be pictured as follows: 
(i) The resonant $\Lambda$ system, Eq.~\eqref{ResonantHamiltonian}, is parametrized in terms of angular variables, Eq.~\eqref{fields-Eulerangles}, and we solve the Euler-Lagrange equations \eqref{EL-nullgrad} for these, with pulse area being the cost function to optimize. 
(ii) The solutions of the explicit equation \eqref{phitrajectory}, which depend on a geometrical parameter ($\dot{\widetilde\phi}_i$), are systematically derived with their corresponding pulse areas in Fig.~\refsubcap{fig_lambdaJandAreavsDphii}{(a)} and \refsubcap{fig_lambdaJandAreavsDphii}{(b)}, and result in the trajectories shown in Figs.~\refsubcap{fig_lambdaJandAreavsDphii}{(c)} and \refsubcap{fig_lambdaJandAreavsDphii}{(d)} for representative values of the parameter $\dot{\widetilde\phi}_i$.
(iii) The robustness of these trajectories is found in Fig.~\ref{fig_robustnessprofile} compared with the optimal non-robust solution \cite{Boscain2002,Boscain2002a}.
(iv) Energy optimization leads to temporal pulse shapes and their respective population dynamics, shown in Fig.~\ref{fig_FieldsAndPopulationsVsTime} for selected values of the parameter $\dot{\widetilde\phi}_i$.
(v) The selected robust and optimal pulse shapes are entirely characterized in Table \ref{tab_loss_vs_area}.

In section \ref{sec_model} we present the model for a resonant three-level system considering a lossy intermediate state; we propose a geometric (angular) parametrization of state, propagator, and loss, while declaring the corresponding boundary conditions appropriate for the $\Lambda$ transfer.
Section \ref{sec_reverse_engineering} contains the parametrization of the fields and definition of the cost functions to optimize, pulse area and energy; the fundamentals of robustness and its manipulation are also discussed.
Section \ref{sec_robust_optimization} gathers the geometric constraints to be enforced, dealing with both: boundary conditions and robustness, and introduces the corresponding Euler-Lagrange equation for the trajectory.
Section \ref{sec_area_optimal_trajectory} presents the results on the robust area-optimization, the optimal trajectories and its defining parameters  and characteristics, irrespective of any specific time parametrization (thus a geometric trajectory in contrast with a temporal dynamics).
Section \ref{sec_energy_optimal_dynamics} shows the Euler-Lagrange equations for the optimization of the state evolution with respect to the generalized pulse energy, leading to a time parametrization that also minimizes the duration of the transfer. 
The temporal shape of the coupling fields, the population dynamics, and the corresponding losses are all discussed in Section \ref{sec_dicussion}, where also the numerical details of the results are gathered.
Finally, conclusions are presented.

Appendixes are included with details for the obtainment of the deviation integrals and the numerical resolution of the trajectory equation.
Some useful geometrical relations for symmetric trajectories and the time-evolution of the angles are also presented as Appendixes.
\section{The model}
\label{sec_model}
We consider a three-level system driven by two resonant fields of Rabi frequencies $\Omega_P(t)$ and $\Omega_S(t)$ for which the Hamiltonian, on the bare states basis $\{\lvert1\rangle,\lvert2\rangle,\lvert3\rangle\}$ and under the rotating wave approximation, is:
\begin{equation}
\label{ResonantHamiltonian}
H_{\Gamma}(t) = \frac{\hbar}{2}\begin{bmatrix}
0&\Omega_P&0\\
\Omega_P&-i\Gamma&\Omega_S\\
0&\Omega_S&0\end{bmatrix},
\end{equation}
and the state of the system is denoted by $\lvert\psi_{\Gamma}(t)\rangle$, solution to the time-dependent Schr\"odinger equation (TDSE) describing the dynamics from the initial to the final times $t_i$ and $t_f$, accounting for dissipation losses of state $\lvert2\rangle$. 
We have considered, as it is standard, that the upper state is lossy through the dissipation rate $\Gamma$.
	
When the dissipation rate is much smaller than the peak Rabi frequency (typically at least 10 times smaller), the total loss of the system during the interaction time $T=t_f-t_i$ is well approximated by 
\begin{equation}
\label{loss}
P_{\text{loss}} \approx \Gamma\int_{t_i}^{t_f}dt\,P_2(t),
\end{equation}
where $P_2=\lvert\langle2\vert\psi_{\Gamma=0}\rangle\rvert^2$ is the population in the excited state in absence of dissipation. 
We will thus consider the dynamics with the lossless Hamiltonian (with $\Gamma=0$) and the expected loss will be taken into account via \eqref{loss}.

Prescribing the desired transfer to be $\lvert\psi(t_i)\rangle\equiv\lvert\psi_i\rangle=\lvert1\rangle\to\lvert\psi(t_f)\rangle\equiv\lvert\psi_f\rangle=\lvert\psi_T\rangle=\pm\lvert3\rangle$, while hoping to maintain a small excited state population (to minimize the loss), we can parametrize the state of the system, solution of the TDSE for the lossless Hamiltonian, $\lvert\psi_{\Gamma=0}\rangle\equiv\lvert\psi\rangle$, as
\begin{align}
\label{single-mode-eigenvector}
\lvert\psi(t)\rangle &= \begin{bmatrix}
\cos\phi\cos\theta\\
i\sin\phi\\
\cos\phi\sin\theta\end{bmatrix},
\end{align}
whose time-dependent angular parametrization must satisfy the boundary conditions:
\begin{subequations}
\label{BoundaryConditons}
\begin{alignat}{3}
\phi_i &= 0&\;\gets\;&\phi(t)&\;\to\phi_f&=0,\\
\label{BoundaryConditons_theta}
\theta_i &= 0&\;\gets\;&\theta(t)&\;\to\theta_f&=\theta_f^\pm=\pm\pi/2.
\end{alignat}
\end{subequations}
The arrows to the right and left indicate the limits when $t\to t_f$ and $t\to t_i$, respectively. 
The sign $\pm$ indicates the two possible options for the terminal $\theta$.
The phase of the target state $\lvert\psi_T\rangle$ is irrelevant for the transfer of population and can be interpreted, and controlled, as a constant carrier-envelope phase difference between the control fields.
Vector \eqref{single-mode-eigenvector} and 
\begin{subequations}
\label{propagatorvectors}
\begin{align}
\label{single-mode-eigenvectorPLUS}
\lvert\psi_+(t)\rangle &= \begin{bmatrix}i\cos\eta\sin\phi\cos\theta-i\sin\eta\sin\theta\\
\cos\eta\cos\phi\\
i\cos\eta\sin\phi\sin\theta+i\sin\eta\cos\theta\end{bmatrix},\\
\label{single-mode-eigenvectorMINUS}
\lvert\psi_-(t)\rangle &= \begin{bmatrix}
-\sin\eta\sin\phi\cos\theta-\cos\eta\sin\theta\\
i\sin\eta\cos\phi\\
-\sin\eta\sin\phi\sin\theta+\cos\eta\cos\theta\end{bmatrix},
\end{align}
\end{subequations}
form a complete dynamical basis and constitute the propagator of the system, $U(t,t_i)=\begin{bmatrix}\lvert\psi\rangle&\lvert\psi_+\rangle&\lvert\psi_-\rangle\end{bmatrix}$, where $\eta(t_i)\equiv\eta_i=0$ is required by definition.
This parametrization for the propagator can also be obtained from the Lewis-Riesenfeld invariant, as in \cite{Laforgue2019}.

The TDSE for the propagator, $H_{\Gamma=0}=i\hbar\dot UU^\dagger$, due particularly to the lack of coupling $\langle 1\vert H_{\Gamma=0}\vert3\rangle$, imposes the condition
\begin{equation}
\label{eta_doteq}
\dot\theta = -\dot\eta\sin\phi
\end{equation}
on the parametrization of the propagator.
From \eqref{single-mode-eigenvector} and \eqref{propagatorvectors}, and by integrating it, we obtain:
\begin{equation}
\label{eta_phi_0}
\int_{t_i}^{t_f}dt\,\dot\eta\sin\phi = -\theta^\pm_f=\mp\frac{\pi}{2},
\end{equation}
which translates the terminal condition $\theta(t_f)\equiv\theta_f$, \eqref{BoundaryConditons_theta}, into a constraint on the time-dependence of $\dot{\eta}$ and $\phi$.

It can be noted that the transient population of the excited state in this representation is given exactly by
\begin{equation}
\label{pop2-formula}
P_2 = \lvert\langle2\vert\psi\rangle\rvert^2=\sin^2\phi,
\end{equation}
and the total time-area of the population on the excited state can be written as:
\begin{equation}
\label{int-pop2-formula}
A_2 = \int_{t_i}^{t_f}dt\,\sin^2\phi\approx\frac{P_\mathrm{loss}}{\Gamma}.
\end{equation}
This area $A_2$ represents thus the loss of the problem normalized by $\Gamma$.
We can see that the presence of the dissipation rate $\Gamma\ne0$ on the upper state induces necessarily a loss to accomplish a $\Lambda$ transfer. 
This is true for any pump and Stokes configuration and shaping, since no loss ($A_2=0$) would require constant $\phi(t)=0$ (for $\Gamma\neq0$), hence constant $\theta(t)=0$ [$\dot\theta(t)=0$ from \eqref{eta_doteq}], and thus no transfer.

As a general strategy to deal with a dissipation rate $\Gamma$, leading to a population loss $P_\mathrm{loss}$, we will solve for the lossless dynamics while striving for low $A_2$'s.
\section{Inverse engineering and robustness}
\label{sec_reverse_engineering}
The parametrization of the TDSE allows one to define the inverse engineering problem: the controls in the Hamiltonian \eqref{ResonantHamiltonian}, with $\Gamma=0$, are expressed in terms of the angles, from the parametrization of the propagator defined by \eqref{single-mode-eigenvector} and \eqref{propagatorvectors}, as
\begin{subequations}
\label{fields-Eulerangles}
\begin{align}
\Omega_P/2 %&= -\dot\theta\cot\phi\sin\theta-\dot\phi\cos\theta\nonumber\\
	&=\dot\eta\cos\phi\sin\theta-\dot\phi\cos\theta,\\
\Omega_S/2 %&= \dot\theta\cot\phi\cos\theta-\dot\phi\sin\theta\nonumber\\
	&=-\dot\eta\cos\phi\cos\theta-\dot\phi\sin\theta.
\end{align}
\end{subequations}
It can be noticed that the values of the time-derivatives at the initial and final time, $\dot\eta_{i,f}$ and $\dot\phi_{i,f}$, give the value of the control fields evaluated at the boundaries, i.e., using \eqref{BoundaryConditons}, 
\begin{subequations}
\begin{alignat}{2}
\Omega_P^i &= -2\dot\phi_i,\qquad   & \Omega_S^i &= -2\dot\eta_i,\\
\Omega_P^f &= \pm2\dot\eta_f,\qquad & \Omega_S^f &= \mp2\dot\phi_f.
\end{alignat}
\end{subequations}
We define the generalized pulse area (referred simply as pulse area from here on) to be
\begin{subequations}
\label{GeneralPulseArea}
\begin{align}
\label{GeneralPulseAreaIntTime}
\area_{t} &\equiv \int_{t_i}^{t_f}dt\,\sqrt{\Omega_P^2+\Omega_S^2}\nonumber\\
		&=2\int_{t_i}^{t_f}dt\,\sqrt{\dot\phi^2+\dot\eta^2\cos^2\phi},
\end{align}
which can be rewritten as an integral in terms of $\eta$ if we assume that $\phi(t)$ can be expressed as a function of $\eta(t)$, i.e.
\begin{align}
\label{GeneralPulseAreaIntEta}
\area 
%&= 2\int_l ds_\eta\,\sqrt{\bigl(\dot{\widetilde\phi}\bigr)^2+\cos^2\widetilde\phi}\nonumber\\
	  &= 2\int_{\eta_i}^{\eta_f}d\eta\,\sgn\dot\eta\sqrt{\bigl(\dot{\widetilde\phi}\bigr)^2+\cos^2\widetilde\phi},
\end{align}
\end{subequations}
where $\widetilde\phi(\eta)\equiv\phi[\eta(t)]$, $\dot{\widetilde\phi}\equiv\partial_\eta\widetilde\phi$, and $\partial_\eta$ is the partial derivative operator with respect to $\eta$.
%and the subscript $l$ under the integral sign denotes a line integral.
Once the sign of $\dot\eta$ is fixed, Eq.~\eqref{GeneralPulseAreaIntEta} will not depend on time but only on the trajectory $\widetilde\phi(\eta)$. Equation \eqref{eta_phi_0} has a similar property.
%We will consider the general situation of a non-monotonic dynamic behavior of $\eta(t)$, requiring to segment the integral and to consider a piecewise function $\widetilde\phi(\eta)$.
The dynamic behavior of $\eta$ can be considered to be monotonic or not, requiring to segment the integral in the latter case and to consider a piecewise function $\widetilde\phi(\eta)$. 
%The smallest pulse area is expected tor a monotonic $\eta$.

The issue of robustness can be dealt with by adding perturbation terms to the Hamiltonian, representing errors or imperfections of the practical implementation.
We consider an error originated by pulse inhomogeneities, taken as identical for both pulses, modeled by the modified Hamiltonian \label{H_eps_pertth} $H_\epsilon=H_{\Gamma=0}+V=(1+\epsilon)H_{\Gamma=0}$, which translates into a deviation on the desired state dynamics and generalized pulse area.
We denote $\lvert\psi_\epsilon(t)\rangle$ as the state of the complete dynamics including the error, solution of the TDSE $i\hbar\partial_t|\psi_{\epsilon}(t)\rangle=H_{\epsilon}\lvert\psi_{\epsilon}(t)\rangle$. 
The single-shot shaped pulse method \cite{Daems2013,Van-Damme2017} allows one to define trajectories, in the dynamical variables space, resistant to errors. 
It can be formulated by a perturbative expansion of $\lvert\psi_{\epsilon}(t_f)\rangle$ with respect to $\epsilon$, $\langle\psi_T\vert\psi_{\epsilon}(t_f)\rangle=1-O_1-O_2-O_3-\cdots$, where $O_n$ denotes the error term of order $n$: $O_n\equiv O(\epsilon^n)$, and $\lvert\psi_T\rangle$ is the target state.

In practice, we 
%can be led to consider piece-wise functions and 
search to attain the optimal solution in terms of certain cost parameter.
For instance, we can define the cost to be the required pulse area to reach the target state, and strive to minimize it; or, we can define the cost to be a specific measure of robustness (e.g., the maximum range of $\epsilon$ for which the target state is reached with under $10^{-4}$ deviation), and maximize it. Here, we will consider both, optimization and robustness, which technically corresponds to \emph{searching the optimal solution with respect to a cost (pulse area, energy, or duration) under the constraint of robustness}.

When we consider both optimization and robustness with respect to the generalized pulse area $\area$ (or identically to both pulse amplitudes for a given time of interaction),
Eqs.~\eqref{eta_phi_0} and \eqref{GeneralPulseArea} show that one can consider the problem in the parameter space formed by the angles $(\eta,\widetilde\phi)$, without invoking a specific time parametrization; thus providing a purely geometric representation of the problem.

In fact, it is known that, in the absence of robustness constraints, minimizing the pulse area \eqref{GeneralPulseArea} is equivalent to minimizing the pulse energy,
\begin{align}
\label{GeneralPulseEnergy}
\energy &= \hbar\int_{t_i}^{t_f}dt\,\bigl(\Omega_P^2+\Omega_S^2\bigr)\nonumber\\
		&= 4\hbar\int_{t_i}^{t_f}dt\,\bigl(\dot\phi^2+\dot\eta^2\cos^2\phi\bigr),
\end{align}
and to minimize the time for a given bound of the pulse amplitudes \cite{Boscain2002} (equivalently, to minimize the pulse amplitudes for a certain pulse duration). We will show that this property still applies for our constrained problem.

%\fbox{the same solution for optimization with respect to the pulse area, pulse energy or pulse duration for a given bound of its amplitude}

A robust optimal transfer of population corresponds to a special trajectory $\widetilde\phi_{\text{opt}}(\eta)$ that, satisfying the boundary conditions \eqref{BoundaryConditons}, minimizes the generalized pulse area \eqref{GeneralPulseAreaIntEta} while attaining robustness up to a certain order.
The construction of the actual time-dependent pulses $\Omega_P$ and $\Omega_S$ from \eqref{fields-Eulerangles} necessitates the use of a specific temporal parametrization, $\eta(t)$, which may be chosen at will (it is inconsequential) for the optimization solely with respect to the pulse area.
On the other hand, optimization with respect to the pulse energy, corresponding to the minimization of Eq.~\eqref{GeneralPulseEnergy}, defines a specific temporal parametrization $\eta_\energy(t)$ for the same optimal trajectory $\widetilde\phi_{\text{opt}}(\eta)$, which also minimizes the pulse duration for a fixed maximum of the pulse amplitudes.
\section{Robust optimal population transfer}
\label{sec_robust_optimization}
For the task of population transfer to a target state $\lvert\psi_T\rangle$, the final global phase is not \emph{a priori} fixed and, since it is irrelevant, its robustness is not cared for. 
The figure of merit up to the third order of robustness reads
\begin{equation}
\label{expansion}
\fidelity = \lvert\langle\psi_T\vert\psi_{\epsilon}(t_f)\rangle\rvert^2=1-\widetilde O_2-\widetilde O_3,
\end{equation}
where the first order is nil (real part of a purely imaginary number which, in this case, is anyway zero), and the second and third orders are
\begin{subequations}
\label{TransferO2O3}
\begin{align}
\label{TransferO2}
\widetilde O_2 &= \biggl\lvert\int_{t_i}^{t_f}dt\,n(t)\biggr\rvert^2+\biggl\lvert\int_{t_i}^{t_f}dt\,p(t)\biggr\rvert^2,\\
\label{TransferO3}
\widetilde O_3 
%&= 2i\int_{t_i}^{t_f}\int_{t_i}^{t'}\int_{t_i}^{t''}\bigl[n(t)r(t')p(t'')\nonumber\\
%&\qquad{}-p(t)r(t')n(t'')\bigr]\,dt''\,dt'\,dt,\\
&= 2i\biggl[\int_{t_i}^{t_f}dt\,n(t)\int_{t_i}^{t_f}dt\!\!\int_{t_i}^tdt'\,r(t)p(t')\nonumber\\
&\qquad{}-\int_{t_i}^{t_f}dt\,p(t)\int_{t_i}^{t_f}dt\!\!\int_{t_i}^tdt'\,r(t)n(t')\biggr],
\end{align}
\end{subequations}
with 
\begin{subequations}
\label{deviation_elements}
\begin{align}
\label{nEqPsiPsiPlus}
n &= \frac{\langle\psi\vert V\vert\psi_+\rangle}{\hbar}=-\dot\eta\sin\eta\sin\phi\cos\phi-\dot\phi\cos\eta,\\  
\label{pEqPsiPsiMinus}
p &= \frac{\langle\psi\vert V\vert\psi_-\rangle}{\hbar}= i( \dot\eta\cos\eta\sin\phi\cos\phi -\dot\phi\sin\eta),\\
r &= \frac{\langle\psi_+\vert V\vert\psi_-\rangle}{\hbar}=-\dot\eta\cos^2\phi.
\end{align}
\end{subequations}
$\epsilon$ was used merely to keep track of the orders of the expansion.
It has been omitted in the above expressions.
We note from \eqref{TransferO2O3} that the only perturbation we need to be concerned about up to third order is $\widetilde O_2$, \eqref{TransferO2}, since \eqref{TransferO3} shows that the third order deviation is null \label{nullO3} for any trajectory $\widetilde\phi(\eta)$ that nullifies the second order [i.e.~the areas under $n(t)$ and $p(t)$].
Some properties used to obtain Eqs.~\eqref{TransferO2O3} are presented in Appendix \ref{AppIntegral}.
\subsection{Lagrangian formulation of the optimization}
The problem of optimal nullification up to the third order can be formulated as a classical optimization problem: finding the trajectory $\widetilde\phi(\eta)$ that minimizes the pulse area \eqref{GeneralPulseAreaIntEta}, which is the action (in the language of Lagrangian mechanics) and integral of a Lagrangian $\lagrangian$,
\begin{align}
\label{intLagrangian}
\area &= 2\int_{\eta_i}^{\eta_f}d\eta\,\sgn\dot\eta\sqrt{\bigl(\dot{\widetilde\phi}\bigr)^2+\cos^2\widetilde\phi}\nonumber\\
	  &\equiv\int_{\eta_i}^{\eta_f}d\eta\,\lagrangian\Bigl(\dot\eta,\widetilde\phi,\dot{\widetilde\phi}\Bigr),
\end{align}
under the constraints $\theta_f^\pm=\pm\pi/2$, from \eqref{eta_phi_0}, and $\widetilde O_2=0$; rewritten for convenience as
\begin{subequations}
\label{ConstraintsPhiOfEta}
\begin{align}
\label{eta_phifct_0}
\xi_0 &= \int_{\eta_i}^{\eta_f}d\eta\,\lvert\sgn\dot\eta\rvert\sin\widetilde\phi\nonumber\\
%	&= \int_{\eta_i}^{\eta_f}d\eta\,\lvert\sgn\dot\eta\rvert\biggl(\sin\widetilde\phi+\frac{\pi}{2\lvert\Delta\eta\rvert}\biggr),\nonumber\\
	&\equiv\int_{\eta_i}^{\eta_f}d\eta\,\varphi_0\bigl(\widetilde\phi\bigr)=-\theta^\pm_f=\mp\frac{\pi}{2},\\
\label{O2n_ct}
\xi_1 &= \int_{\eta_i}^{\eta_f}d\eta\,\lvert\sgn\dot\eta\rvert\Bigl(\dot{\widetilde{\phi}}\cos\eta+\sin\eta\sin\widetilde\phi\cos\widetilde\phi\Bigr)\nonumber\\
	&\equiv\int_{\eta_i}^{\eta_f}d\eta\,\varphi_1\bigl(\eta,\widetilde\phi,\dot{\widetilde\phi}\bigr)=0,\\
\label{O2p_ct}
\xi_2 &= \int_{\eta_i}^{\eta_f}d\eta\,\lvert\sgn\dot\eta\rvert\Bigl(\dot{\widetilde\phi}\sin\eta-\cos\eta\sin\widetilde\phi\cos\widetilde\phi\Bigr)\nonumber\\
	&\equiv\int_{\eta_i}^{\eta_f}d\eta\,\varphi_2\bigl(\eta,\widetilde\phi,\dot{\widetilde\phi}\bigr)=0,
\end{align}
\end{subequations} 
while satisfying the boundary conditions, for which the initial state is characterized by the angles $(\theta_i=0,\phi_i=0,\eta_i=0)$ and the target (final) state by $(\theta_f^\pm=\pm\pi/2,\phi_f=0,\eta_f$). 
The factor $\lvert\sgn\dot\eta\rvert$ was added only as a reminder that we are dealing with a piecewise function $\widetilde\phi(\eta)$, where the interval of integration must be split each time $\dot\eta$ has a sign change. A way to detect such change of sign can be achieved geometrically during the determination of the trajectory $\widetilde\phi(\eta)$ from the initial condition (starting with a given sign of $\dot\eta$). The change of sign can occur at a point $\eta_0$ only when $\vert \dot{\widetilde\phi}(\eta_0)\vert\to\infty$. We will see that this does not happen in our problem, and that a monotonic $\eta(t)$ can be considered.

In this representation it is thus relevant to consider the trajectories $\widetilde\phi(\eta)$, constrained by the conditions \eqref{ConstraintsPhiOfEta}, in the parameter space $(\eta,\widetilde\phi)$.
\subsection{Derivation of the trajectory \texorpdfstring{$\widetilde\phi(\eta)$}{phi(eta)}}
We consider the representation of the trajectory $\widetilde\phi(\eta)$.
Robust optimal control can be attained by solving the Euler-Lagrange equations and using the Lagrange multiplier method to account for the constraints.
\label{taskcompPT} The task of complete population transfer, for the lossless system, is part of the constraints to be imposed, which is equivalent to enforcing the boundary conditions.
In this context, complete population transfer refers to satisfying the boundary conditions \eqref{BoundaryConditons}, leaving the loss to be estimated \emph{a posteriori} via \eqref{loss}.

The optimal trajectory $\widetilde\phi(\eta)$ is a solution of
\begin{equation}
\label{EL-nullgrad}
\grad\area+\sum_{j=0}^{2}\lambda_j\grad\xi_j=0,
\end{equation}
where $\lambda_j$ ($j=0,1,2$) is the Lagrangian multiplier associated to each one of the three constraints, and the gradients,
\begin{subequations}
\begin{align}
\grad\area &= \frac{\partial\lagrangian}{\partial\widetilde\phi}-\frac{d}{d\eta}\Biggl(\frac{\partial\lagrangian}{\partial \dot{\widetilde\phi}}\Biggr),\\
\grad\xi_j &= \frac{\partial\varphi_j}{\partial\widetilde\phi}-\frac{d}{d\eta}\Biggl(\frac{\partial\varphi_j}{\partial \dot{\widetilde\phi}}\Biggr),
\end{align}
\end{subequations}
are defined according to the Euler-Lagrange equations.

We proceed to obtain the differential equation for the trajectory $\widetilde\phi(\eta)$ from:
\begin{align}
&\frac{\partial\lagrangian}{\partial\widetilde\phi}-\frac{d}{d\eta}\Biggl(\frac{\partial\lagrangian}{\partial\dot{\widetilde\phi}}\Biggr)\nonumber\\
&\qquad{}+\sum_{j=0}^2\lambda_j\Biggl[\frac{\partial\varphi_j}{\partial\widetilde\phi}-\frac{d}{d\eta}\Biggl(\frac{\partial\varphi_j}{\partial\dot{\widetilde\phi}}\Biggr)\Biggr]=0,
\end{align}
which leads, after simplification by $2\cos^2\widetilde\phi$, to
\begin{align}
\label{EulerLagrangePhiOfEta}
&-\sgn\dot{\eta}\frac{\ddot{\widetilde\phi}+\Bigl[2\bigl(\dot{\widetilde\phi}\bigr)^2+\cos^2\widetilde\phi\Bigr]\tan\widetilde\phi}{\Bigl[\bigl(\dot{\widetilde\phi}\bigr)^2+\cos^2\widetilde\phi\Bigr]^{3/2}}\nonumber\\
&\qquad{}+\lvert\sgn\dot\eta\rvert\bigl(\lambda_0\sec\widetilde\phi+\lambda_1\sin\eta-\lambda_2\cos\eta\bigr)=0.
\end{align}
Note that we have redefined $\lambda_0/2$ as $\lambda_0$ without loss of generality.

Solving \eqref{EulerLagrangePhiOfEta} means to find a trajectory $\widetilde\phi(\eta)$ with the $\lambda_j$'s as free parameters to be set to satisfy the constraints $\eqref{ConstraintsPhiOfEta}$.
We can solve this numerically assuming a monotonic behavior for $\eta(t)$, i.e.
\begin{align}
\label{phitrajectory}
&\mp\frac{\ddot{\widetilde\phi}{}^\pm+\Bigl[2\bigl(\dot{\widetilde\phi}{}^\pm\bigr)^2+\cos^2\widetilde\phi^\pm\Bigr]\tan\widetilde\phi^\pm}{\Bigl[\bigl(\dot{\widetilde\phi}{}^\pm\bigr)^2+\cos^2\widetilde\phi^\pm\Bigr]^{3/2}}\nonumber\\
&\qquad{}+\lambda_0\sec\widetilde\phi^\pm+\lambda_1\sin\eta-\lambda_2\cos\eta=0,
\end{align}
where we have used $\widetilde{\phi}^\pm\equiv\widetilde\phi_{\sgn\dot\eta=\pm1}$.
\section{Robust area-optimal trajectory \texorpdfstring{$\widetilde\phi(\eta)$}{phi(eta)}}
\label{sec_area_optimal_trajectory}
\begin{figure}
\hypertarget{fig_lambdaJandAreavsDphii}{}%
\centering
\includegraphics[scale=1]{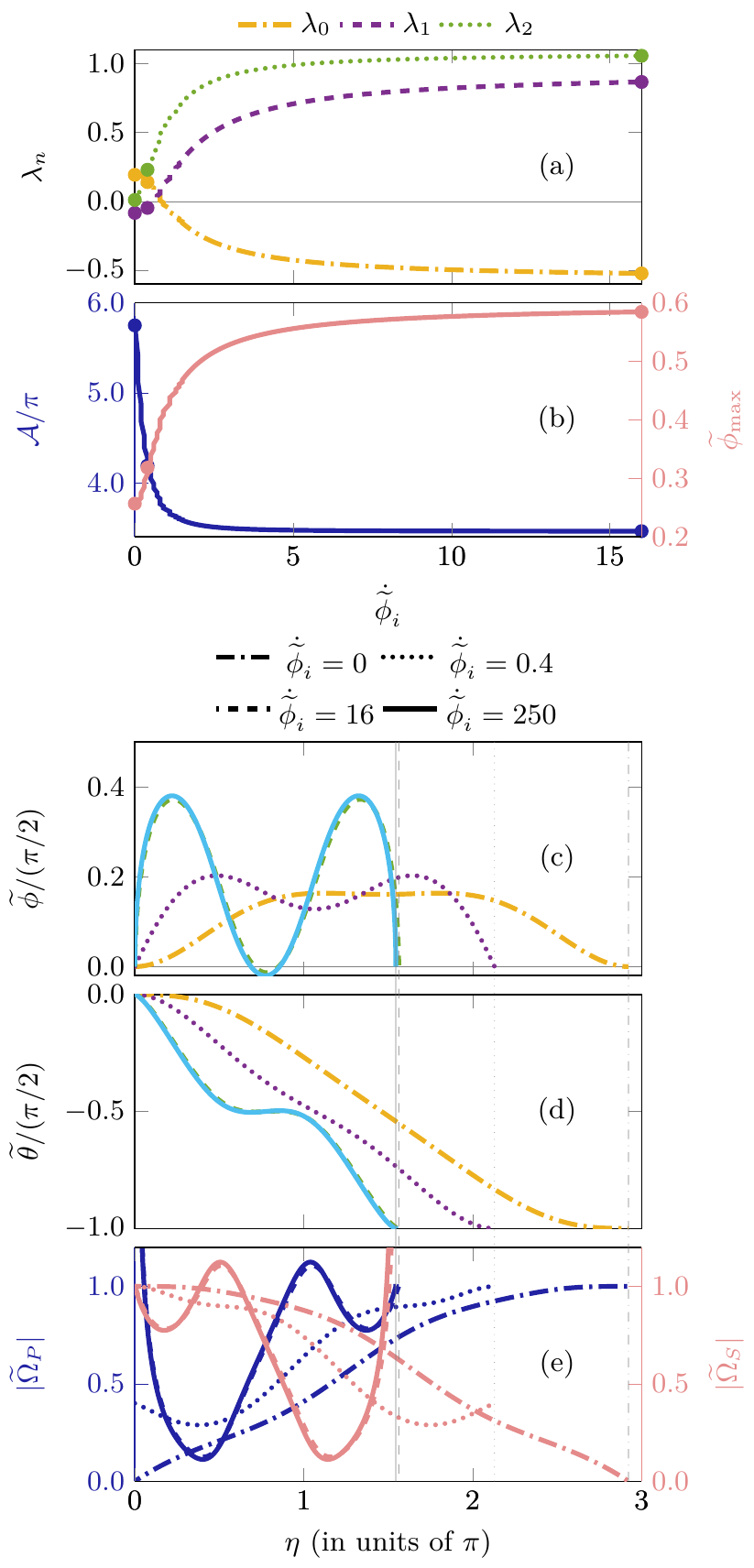}
\caption{\label{fig_lambdaJandAreavsDphii}Area-optimal solutions vs $\dot{\widetilde\phi}_i$ regarding (a) the $\lambda_j$'s and (b) the pulse area and maximum value of $\widetilde\phi$. 
Trajectories $\widetilde\phi(\eta)$ and $\widetilde\theta(\eta)$, for selected extrema of the area minimization problem, are shown in (c) and (d).
Respective Pump and Stokes fields, dynamically scaled by $2\dot\eta$, are shown in (e) vs $\eta$.
The parameters defining the highlighted extrema are summarized in Table \ref{tab_loss_vs_area}.
The line style in the legend applies to all plots irrespective of the line color.
The thin vertical gray lines are located at the $\eta_f$ corresponding to each trajectory.
Thin horizontal gray lines mark a zero.}
\end{figure}

We determine the solution for the robust area-optimal trajectory via the numerical implementation of \eqref{phitrajectory} into an ordinary differential equations solver [see system \eqref{numerical_implementation} in Appendix \ref{AppNumImpl}] and use its solution, in terms of the parameters $\bigl(\dot{\widetilde\phi}_i,\lambda_0,\lambda_1,\lambda_2\bigr)$, for a subsequent nonlinear equations solver that seeks to satisfy the four trajectory constraints by searching in that four-parameter space.
It turns out that for each value of $\dot{\widetilde\phi}_i$ there is a trajectory solution to the Euler-Lagrange equations satisfying the imposed  constraints.
The parameters solution of this system are presented in Fig.~\refsubcap{fig_lambdaJandAreavsDphii}{(a)} for values of $0\leq\dot{\widetilde\phi}_i\leq16$.

The corresponding generalized pulse areas and the value of \label{phitildenotphi} $\widetilde\phi$ at the summit of the respective trajectories (related to the normalized loss $A_2$) are shown in Fig.~\refsubcap{fig_lambdaJandAreavsDphii}{(b)}.

The maximum and minimum generalized pulse areas on the plot, Fig.~\refsubcap{fig_lambdaJandAreavsDphii}{(b)}, are, respectively, $\area_{\textrm{max}}=5.7498\pi$ and $\area_{\textrm{min}}=3.4608\pi$, while the corresponding minimum and maximum values of $\phi$ (inversely related to the area) are $\phi_{\textrm{min}}=0.2566$ and $\phi_{\textrm{max}}=0.5893$.
If to extend the plot to a large value, much beyond the point were significative change occurs on the trajectory, e.g., $\dot{\widetilde\phi}_i=250$, we would obtain $\area=3.4603\pi$ and $\widetilde\phi_{\textrm{max}}=0.5975$.

The trajectories represented by each set of points corresponding to a single value of $\dot{\widetilde\phi}_i$ are the extrema of the optimization problem, candidates to be an optimal solution.
The optimum trajectory is obtained among all these solutions for the one corresponding to the minimum generalized pulse area: $\area_{\textrm{min}}\approx3.4603\pi$, associated to $|\dot{\widetilde\phi}_i|\to\infty$ and to a normalized loss $A_2\approx0.1291$.
However, we highlight that all the other extremal solutions, featuring larger pulse areas, represent physical optimal and robust solutions, but with corresponding lower normalized losses.

\label{wenoticesym}We notice that these solutions, defined by Fig.~\refsubcap{fig_lambdaJandAreavsDphii}{(a)} and extensible to $\dot{\widetilde\phi}_i\to\infty$, are symmetrically mirrored (with identical pulse areas) for $\dot{\widetilde\phi}_i\leq0$, with a sign change in the $\lambda_j$'s (and $\phi_{\textrm{max}}$), thus the sign-changed alternative solution at $\dot{\widetilde\phi}_i=0$ was omitted for clarity.
These alternative trajectories, namely with sign-changed $\widetilde\phi$ and $\widetilde\theta$, produce sign-changed pumps and final states, $\Omega_P$ and $\lvert\psi_f\rangle$, and do not differ in any other way; hence, we limit our analysis to $\dot{\widetilde\phi}_i\geq0$.

Although it is far from being actually adiabatic (much lower areas compared to usual adiabatic requirements), the behavior of this family of solutions is reminiscent of adiabatic solutions in terms of correlation between pulse area and normalized loss: higher the invested pulse area, lower the maximum of the transient excited state population and the corresponding normalized loss (see Fig. \ref{fig_A2vsArea}).
%However, this is caused by adiabaticity only for much larger areas and not on such a smooth manner (lacking the local defects observed on adiabatic methods due to non-adiabatic dynamics).

It is worthy of mention that Eq.~\eqref{phitrajectory} was solved [i.e.~the system \eqref{numerical_implementation} was integrated] without demanding the symmetry of the trajectory, see Appendix \ref{Appsymmtraj}.
Each trajectory was computed from $\eta=0$ to a large $\eta_{\textrm{max}}$ (typically $\eta_{\textrm{max}}=3\pi$).
Then, the solutions were obtained by taking $\phi_f$ to be a point where the trajectory crossed the $\phi=0$ line boundary (thus truncating there the trajectory).

Having left the symmetry (or lack of it) of the trajectories to be decided by the solution of the dynamical system and satisfaction of the constraints, we obtained symmetric (of even parity) trajectories with $\eta_f$ matching (one of) the expected values \eqref{cosetaf} naturally, as it can be seen in Figs.~\refsubcap{fig_lambdaJandAreavsDphii}{(c)}--\refsubcap{fig_lambdaJandAreavsDphii}{(e)}.

The values of $\eta_f$ are lower for higher maximum state populations, but the resultant areas are lower; trajectories and pulses are then shorter (in $\eta$), leading, presumably, to the optimal time: the fastest way to go from $\eta_i\equiv0$ to $\eta_f$, which would naturally be faster for lower values of $\eta_f$.

We notice that all solutions for $\dot{\widetilde\phi}_i\leq5.5$ were obtained by choosing $\phi_f$ as the first zero-crossing, while for $\dot{\widetilde\phi}_i>5.5$ we had to truncate at the third one.

\label{Otherfams}Other families of solutions, families of extrema of the optimization problem, were found for larger boundaries of the integration ($\eta_{\text{max}}>3\pi$) and for second- and third-zero--crossings, some of them even displaying asymmetric trajectories.
However, all of them presented larger areas to the family in Fig.~\ref{fig_lambdaJandAreavsDphii}, thus they are irrelevant to the problem of optimization.
For example, the asymmetric trajectories corresponding to the second--zero-crossing family of solutions present areas of $5.715\pi\leq\area\leq5.750\pi$.
Meanwhile, the next immediate family of trajectories (third--zero-crossing) exhibit about three times the area and value of $\eta_f$, in the vicinity of $\dot{\widetilde\phi}_i=0$, of the optimal family.

The extremum trajectories $\widetilde\phi(\eta)$ in Fig.~\refsubcap{fig_lambdaJandAreavsDphii}{(c)} display a double peak structure, although it is only slight for the the largest-area extremum, $\dot{\widetilde\phi}_i=0$.
The well between the persistent positive peaks becomes a negative peak (though of much smaller magnitude) for the optimal trajectory.
By comparison between the shown trajectories with the largest $\dot{\widetilde\phi}_i$ it is clear that the system geometric evolution is optimal at infinity, but values in the order of the tens already describe it well.

The evolution of the mixing angle $\widetilde\theta(\eta)$ behaves as a symmetrical two-step process, two identical consecutive evolutions $0\to\pi/4\to\pi/2$; however, it can not be regarded as twice a robust half-transfer, since, although we can take the extrema whose half-point is nil as its endpoint, $\phi^{\textrm{half}}_f=\phi_m=0$, this half-transfer is not robust [since it does not satisfy \eqref{ConstraintsPhiOfEta} for $\eta^\textrm{half}_f=\eta_m$].

Having treated the dynamics as a geometric trajectory $\widetilde\phi(\eta)$, we are left only with the dynamically-scaled fields
\begin{subequations}
\label{dynscaledfields}
\begin{align}
\widetilde \Omega_P(\eta) &\equiv \Omega_P/(2\dot\eta)= \cos\widetilde\phi\sin\widetilde\theta-\dot{\widetilde\phi}\cos\widetilde\theta,\\
\widetilde \Omega_S(\eta) &\equiv \Omega_S/(2\dot\eta)=-\cos\widetilde\phi\cos\widetilde\theta-\dot{\widetilde\phi}\sin\widetilde\theta,
\end{align}
\end{subequations}
to picture the control fields in terms of $\eta$.
These equations show that the initial and final values of these parametrized ratios are:
\begin{subequations}
\label{scaledfieldsboundaries}
\begin{alignat}{2}
{\widetilde\Omega_P}^i &= -\dot{\widetilde\phi}_i, & \qquad {\widetilde\Omega_P}^f &= \sgn\theta_f,\\
{\widetilde\Omega_S}^i &= -1, & \qquad {\widetilde\Omega_S}^f &= -\sgn\theta_f\dot{\widetilde\phi}_f.
\end{alignat}
\end{subequations}
The absolute value of the dynamically-scaled control fields $\widetilde\Omega_P$ and $\widetilde\Omega_S$ is shown in Fig.~\refsubcap{fig_lambdaJandAreavsDphii}{(e)}, evidencing the boundary Eqs.~\eqref{scaledfieldsboundaries}.
Although the actual time-dependent control fields will be described only after a time-parametrization of $\eta(t)$ is decided, we can already note that the couplings present a marked counter-intuitive ordering.
This is obvious for the largest-area extrema, but it is still mostly true for the optimum with the exception of the spikes near the boundaries of the trajectory.

The ideal robust optimum solution would demand infinite scaled amplitudes at one of the boundaries for each field, but this does not require the actual pulses to have infinite magnitudes at any point in time.
The fields as functions of time may be made indeed finite with a proper choice of the time parametrization, i.e.~of $\eta(t)$.

The quantities $\bigl(\widetilde\Omega_P,\widetilde\Omega_S\bigr)$ are sufficient to solve the TDSE parametrized in terms of $\eta$, with $\widetilde H_\epsilon(\eta)\equiv H_\epsilon(t)/\dot\eta$,
\begin{align}
%i\hbar\partial_t\lvert\psi_{\epsilon}(t)\rangle &= H_{\epsilon}(t)\lvert\psi_{\epsilon}(t)\rangle,\\\nonumber
i\hbar\partial_\eta\lvert\widetilde\psi_{\epsilon}(\eta)\rangle &= \widetilde H_{\epsilon}(\eta)\lvert\widetilde\psi_{\epsilon}(\eta)\rangle,
\end{align}
which can be reparametrized back to time, to observe populations actual temporal-dynamics, by simply providing the time-dependence of $\eta(t)$.

The non-robust optimal exact $\Lambda$ transfer has been derived in \cite{Boscain2002} and their corresponding analytical pulse shapes for pulse-energy optimization were given in \cite{Boscain2002a}, producing the coupling fields:
\begin{subequations}
\label{cos-sin}
\begin{align}
\Omega_P &= \frac{\sqrt{3}\pi}{T}\cos\Bigl[\frac{\pi(t-t_i)}{2T}\Bigr],\\
\Omega_S &= \frac{\sqrt{3}\pi}{T}\sin\Bigl[\frac{\pi(t-t_i)}{2T}\Bigr],
\end{align}
\end{subequations}
where $T$ is the pulse duration.
These coupling fields of generalized area $\sqrt3\pi$ are the equivalent of the $\pi$-pulse Rabi solution, the diabatic solution by excellence, for the three-level $\Lambda$ system.
It exhibits the minimum area and energy necessary to perform the complete population transfer $\lvert1\rangle\to\lvert3\rangle$, while the pulse duration fixes the cap on the field amplitudes (equivalently we may fix the pulse amplitudes and extract the minimum time).
Hence, it is the ideal benchmark to test the gained robustness of our optimal robust solution; just as the $\pi$-pulse would be used to compare with population-inversion schemes in a two-level system \cite{Daems2013,Dridi2020}.
This optimal solution, as it differentiates from our robust optimal results in the fact that it doesn't satisfy the robustness constraints, will be referred to as the unconstrained optimal, or simply optimal, solution.
Therefore, another use of this basis for comparison is to use it to understand what is the minimum energy required to gain or acquire a certain order of robustness.

\begin{figure}
\centering
\includegraphics[scale=1]{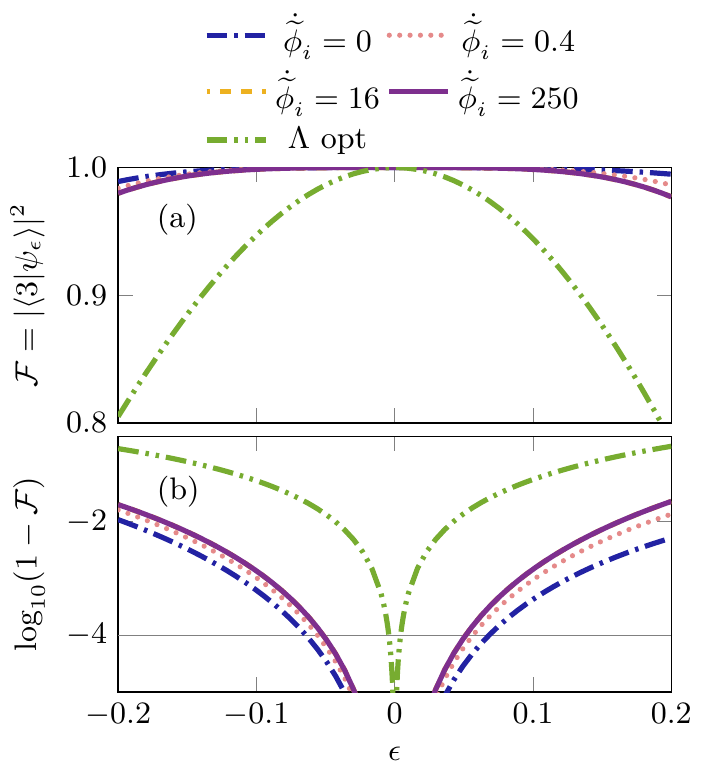}
\caption{Fidelity (a) and logarithm base 10 of the infidelity (b), for the selected local optima highlighted in Fig.~\ref{fig_lambdaJandAreavsDphii} with numeric data summarized in Table \ref{tab_loss_vs_area} and for the $\Lambda$ transfer optimized with no constraints of robustness, with respect to the pulse area scaling error $\epsilon$.
The thin horizontal gray line denotes the ultrahigh-fidelity benchmark of $10^{-4}$ infidelity.}\label{fig_robustnessprofile}
\end{figure}

Solving the TDSE taking into account pulse area scaling error up to $\pm20\%$, we obtain the robustness profile of the population transfer fidelity and base 10 logarithm of the infidelity, presented in Figs.~\ref{fig_robustnessprofile}(a) and \ref{fig_robustnessprofile}(b), respectively.
The profile is slightly broader for larger-area extrema, but all extrema are much more robust than the unconstrained pulse area-optimal solution \eqref{cos-sin}.
Particularly, the fidelity profile for the optimal extrema is symmetric around the unperturbed condition, while the extrema with larger areas present an advantageous slanted profile towards the positive area-scaling deviations ($\epsilon>0$).

\label{thesymmadiabprev}The symmetry of the optimal solutions is a remarkable feature which is certainly not shared with adiabatic processes (or any other scheme for that matter).
For the robustness profile, an increase on energy (equivalently, area) in the adiabatic regime is indeed not equivalent to a decrease.
This is shown in ref.~\cite{Laforgue2019}, where the robustness profile is determined for an exact $\Lambda$ transfer: the slanting of it is evident for larger areas where adiabaticity starts to prevail.
Understanding adiabaticity as the general behavior or tendency of the evolution approaching the dynamics at the adiabatic limit.
We can note that the optimal with no constraints of robustness displays a strictly symmetrical robustness profile, which is a shared characteristic with our optimal robust solution.
%It would be interesting to examine if this observation could be used as a qualitative indicator to differentiate adiabatic from non-adiabatic solutions, or even to characterize their degree of adiabaticity; considering that both optimal solutions, the unconstrained and robust ones, present symmetric profiles while the local minima part of the family of robust solutions present the asymmetries observed for protocols with a strong adiabatic component.

The effect of investing about one time more the area of the optimal pulse ($\area_\textrm{RobOpt}\approx 2\area_\textrm{opt}$) is remarkable: almost 13 times gain in the width of the robustness between $\epsilon=0$ and the closest point where infidelity goes above $10^{-4}$, i.e.~$\Delta\epsilon_\textrm{RobOpt}\approx0.051$ while $\Delta\epsilon_\textrm{opt}\approx0.004$.
\label{itisworthnoting}It is worth noting that the unconstrained optimal pulses are intuitively ordered while our robust optimal pulses feature an overall counter-intuitive ordering.
As observed in Fig.~\refsubcap{fig_lambdaJandAreavsDphii}{(e)}, the largest-area robust extrema (for $\dot{\widetilde\phi}_i=0$) presents the most simple and clearly counter-intuitively--ordered pulse pair.
Meanwhile, the actually-minimal--area extrema (global optimum), corresponding to $\dot{\widetilde\phi}_i\to\infty$, exhibits counter-intuitive behavior at most instances of the ``dynamics'' (actually, geometric evolution along $\eta$), i.e.~except near the beginning and the end of the evolution.
\section{Robust energy-optimal dynamics \texorpdfstring{$\eta(t)$}{eta(t)}}
\label{sec_energy_optimal_dynamics}
The time dependence of the area-optimal geometrical trajectory is free to be chosen.
However, it is most interesting to consider optimality with respect to the pulse energy, which would also satisfy optimality with respect to time for a certain maximum amplitude, as shown below.
We can do this with the Euler-Lagrange equations with constraints, as we have done for the area optimization, but using the energy definition
\begin{align}
\label{PulseEnergyIntEta}
\energy &= \hbar\int_{t_i}^{t_f}dt\,\bigl(\Omega_P^2+\Omega_S^2\bigr)=4\hbar\int_{t_i}^{t_f}dt\,\bigl(\dot{\phi}^2+\dot\eta^2\cos^2\phi\bigr),\nonumber\\
&\equiv \int_{t_i}^{t_f}dt\, \lagrangian_\energy (\phi,\dot\eta,\dot\phi),
\end{align}
as the cost to be minimized.
In this case, the time-representation of the constraints \eqref{ConstraintsPhiOfEta} writes:
\begin{subequations}
\label{ConstraintsPhiOfEta_t}
\begin{align}
\label{eta_phifct_0_t}
\xi_{t0} &= \int_{t_i}^{t_f}dt\,\dot\eta\sin\phi\equiv \int_{t_i}^{t_f}dt\,\varphi_{t0}(\phi,\dot\eta)\nonumber\\
	&=-\theta^\pm_f=\mp\frac{\pi}{2},\\
\xi_{t1} &= \int_{t_i}^{t_f}dt\,(\dot{\phi}\cos\eta+\dot{\eta}\sin\eta\sin\phi\cos\phi)\nonumber\\
&\equiv\int_{t_i}^{t_f}dt\,\varphi_{t1}(\eta,\phi,\dot\eta,\dot{\phi})=0,\\
\xi_{t2} &= \int_{t_i}^{t_f}dt\,(\dot{\phi}\sin\eta-\dot\eta\cos\eta\sin\phi\cos\phi)\nonumber\\
	&\equiv\int_{t_i}^{t_f}dt\,\varphi_{t2}(\eta,\phi,\dot\eta,\dot{\phi})=0,
\end{align}
\end{subequations}
which are satisfied, regardless of the time dependence of $\eta$, by the trajectories $\widetilde\phi(\eta)$ from \eqref{EulerLagrangePhiOfEta} with their appropriate choices of the $\lambda_j$'s.

The dynamics $\{\theta(t),\eta(t),\phi(t)\}$ can be formulated as the solution of the optimal problem, as the evolution satisfying the Euler-Lagrange equations
\begin{equation}
\label{LagrangeMultdelta}
\grad\energy+\mu_0\grad\psi_{t0}+\mu_1\grad\psi_{t1}+\mu_2\grad\psi_{t2}=0,
\end{equation}
where the $\mu_j$'s ($j=0,1,2$) are the Lagrangian multipliers associated to the constraints, and the gradients are
\begin{subequations}
\begin{align}
\grad\energy &= \begin{bmatrix}
\frac{\partial\lagrangian_\energy}{\partial\eta}-\frac{d}{dt}\Bigl(\frac{\partial\lagrangian_\energy}{\partial\dot{\eta}}\Bigr)\\
\frac{\partial\lagrangian_\energy}{\partial\phi}-\frac{d}{dt}\Bigl(\frac{\partial\lagrangian_\energy}{\partial \dot{\phi}}\Bigr)
\end{bmatrix},\\
\grad\xi_{tj} &= \begin{bmatrix}
\frac{\partial\varphi_{tj}}{\partial\eta}-\frac{d}{dt}\Bigl(\frac{\partial\varphi_{tj}}{\partial\dot{\eta}}\Bigr)\\
\frac{\partial\varphi_{tj}}{\partial\phi}-\frac{d}{dt}\Bigl(\frac{\partial\varphi_{tj}}{\partial\dot{\phi}}\Bigr)
\end{bmatrix}.
\end{align}
\end{subequations}%
The Euler-Lagrange equations lead to
\begin{subequations}
\begin{align}
\label{EL1}
0 &= -\frac{d}{dt}\biggl(\frac{\partial\lagrangian_\energy}{\partial\dot{\eta}}\biggr)+\sum_{j=0}^2\mu_j\biggl[\frac{\partial\varphi_{tj}}{\partial\eta}-\frac{d}{dt}\biggl(\frac{\partial\varphi_{tj}}{\partial\dot{\eta}}\biggr)\biggr],\\
%&=-\frac{d}{dt}\bigl(8\dot\eta\cos^2\phi\bigr)-\mu_0\frac{d}{dt}\bigl(\sin\phi\bigr)\nonumber\\
%&\qquad{}+\mu_1\biggl[-\dot{\phi}\sin\eta+\dot{\eta}\cos\eta\sin\phi\cos\phi\nonumber\\
%&\qquad{}-\frac{d}{dt}\biggl(\sin\eta\sin\phi\cos\phi\biggr)\biggr]+\mu_2\biggl[\dot{\phi}\cos\eta\nonumber\\
%&\qquad{}+\dot\eta\sin\eta\sin\phi\cos\phi-\frac{d}{dt}\biggl(-\cos\eta\sin\phi\cos\phi\biggr)\biggr]\nonumber\\
%&=-8\bigl(\ddot\eta\cos^2\phi-2\dot\eta\dot\phi\cos\phi\sin\phi\bigr)-\mu_0\dot\phi\cos\phi\nonumber\\
%&\qquad{}+\mu_1\biggl[-\dot{\phi}\sin\eta+\dot{\eta}\cos\eta\sin\phi\cos\phi\nonumber\\
%&\qquad{}-\biggl(\dot\eta\cos\eta\sin\phi\cos\phi+\dot\phi\sin\eta\cos2\phi\biggr)\biggr]\nonumber\\
%&\qquad{}+\mu_2\biggl[\dot{\phi}\cos\eta+\dot\eta\sin\eta\sin\phi\cos\phi\nonumber\\
%&\qquad{}-\biggl(\dot\eta\sin\eta\sin\phi\cos\phi-\dot\phi\cos\eta\cos2\phi\biggr)\biggr]\nonumber\\
%&=-8\cos^2\phi\bigl(\ddot\eta-2\dot\eta\dot\phi\tan\phi\bigr)-\dot\phi\mu_0\cos\phi\nonumber\\
%&\qquad{}-2\dot{\phi}\mu_1\sin\eta\cos^2\phi+2\dot{\phi}\mu_2\cos\eta\cos^2\phi\nonumber\\
%&=-2\cos^2\phi\biggl[4\bigl(\ddot\eta-2\dot\eta\dot\phi\tan\phi\bigr)+\dot\phi\biggl(\frac{\mu_0}{2}\sec\phi\nonumber\\
%&\qquad{}+\mu_1\sin\eta-\mu_2\cos\eta\biggr)\biggr],\\
\label{EL2}
0 &= \frac{\partial\lagrangian_\energy}{\partial\phi}-\frac{d}{dt}\biggl(\frac{\partial\lagrangian_\energy}{\partial\dot{\phi}}\biggr)+\sum_{i=0}^2\mu_i\biggl[\frac{\partial\varphi_{tj}}{\partial\phi}\nonumber\\
	&\qquad{}-\frac{d}{dt}\biggl(\frac{\partial\varphi_{tj}}{\partial\dot{\phi}}\biggr)\biggr],%\nonumber\\
%&= -8\dot\eta^2\cos\phi\sin\phi-8\ddot\phi+\dot\eta\mu_0\cos\phi\nonumber\\
%&\qquad{}+\mu_1\biggl[\dot\eta\sin\eta\cos2\phi-\frac{d}{dt}\biggl(\cos\eta\biggr)\biggr]\nonumber\\
%&\qquad{}+\mu_2\biggl[-\dot\eta\cos\eta\cos2\phi-\frac{d}{dt}\biggl(\sin\eta\biggr)\biggr]\nonumber\\
%&= -8\dot\eta^2\cos\phi\sin\phi-8\ddot\phi+\dot\eta\mu_0\cos\phi+\mu_1\biggl[\dot\eta\sin\eta\cos2\phi\nonumber\\
%&\qquad{}+\dot\eta\sin\eta\biggr]+\mu_2\biggl[-\dot\eta\cos\eta\cos2\phi-\dot\eta\cos\eta\biggr]\nonumber\\
%&= -8\dot\eta^2\cos\phi\sin\phi-8\ddot\phi+\dot\eta\mu_0\cos\phi+2\dot\eta\mu_1\sin\eta\cos^2\phi\nonumber\\
%&\qquad{}-2\dot\eta\mu_2\cos\eta\cos^2\phi\nonumber\\
%&= -2\biggl[4\ddot\phi+4\dot\eta^2\sin\phi\cos\phi-\dot\eta\cos^2\phi\biggl(\frac{\mu_0}{2}\sec\phi\nonumber\\
%&\qquad{}+\mu_1\sin\eta-\mu_2\cos\eta\biggr)\biggr]\nonumber\\
%&=4\ddot\phi+4\dot\eta^2\sin\phi\cos\phi-\dot\eta\cos^2\phi\biggl(\frac{\mu_0}{2}\sec\phi\nonumber\\
%&\qquad{}+\mu_1\sin\eta-\mu_2\cos\eta\biggr),
\end{align}
\end{subequations}
which are, for $\cos\phi\ne0$,
\begin{subequations}
\begin{align}
\label{EL1_}
\ddot\eta &= \frac{\dot\phi}{4}\biggl[8\dot\eta\tan\phi-\frac1\hbar\biggl(\frac{\mu_0}{2}\sec\phi+\mu_1\sin\eta\nonumber\\
  		  &\qquad{}-\mu_2\cos\eta\biggr)\biggr],\\
\label{EL2_}
\ddot\phi &= -\frac{\dot\eta}{4}\cos^2\phi\biggl[4\dot\eta\tan\phi-\frac1\hbar\biggl(\frac{\mu_0}{2}\sec\phi+\mu_1\sin\eta\nonumber\\
		  &\qquad{}-\mu_2\cos\eta\biggr)\biggr].
\end{align}
\end{subequations}
We can combine these, by performing $\dot\eta\cos^2\phi\,\eqref{EL1_}+\dot\phi\,\eqref{EL2_}$, to obtain a relation not encumbered by the Lagrange multipliers:
\begin{align}
%\ddot\eta\dot\eta\cos^2\phi+\ddot\phi\dot\phi 
%&= \dot\eta\cos^2\phi\biggl[\frac{\dot\phi}{4}\biggl(8\dot\eta\tan\phi-\frac{\lambda_0}{2}\sec\phi-\lambda_1\sin\eta\nonumber\\
%&\qquad{}+\lambda_2\cos\eta\biggr)\biggr]+\dot\phi\biggl[-\frac{\dot\eta}{4}\cos^2\phi\biggl(4\dot\eta\tan\phi-\frac{\lambda_0}{2}\sec\phi\nonumber\\
%&\qquad{}-\lambda_1\sin\eta+\lambda_2\cos\eta\biggr)\biggr]\nonumber\\
%&= \frac{\dot\phi\dot\eta}{4}\cos^2\phi\biggl[\biggl(8\dot\eta\tan\phi-\frac{\lambda_0}{2}\sec\phi-\lambda_1\sin\eta\nonumber\\
%&\qquad{}+\lambda_2\cos\eta\biggr)-\biggl(4\dot\eta\tan\phi-\frac{\lambda_0}{2}\sec\phi\nonumber\\
%&\qquad{}-\lambda_1\sin\eta+\lambda_2\cos\eta\biggr)\biggr]\nonumber\\
%&= \dot\phi\dot\eta^2\cos\phi\sin\phi,\\
\frac{d}{dt}\Bigl(\dot\phi^2+\dot\eta^2\cos^2\phi\Bigr) &= \frac{d}{dt}\biggl\{\dot\eta^2\Bigl[\Bigl(\dot{\widetilde\phi}\Bigr)^2+\cos^2\widetilde\phi\Bigr]\biggr\}\nonumber\\
&= \frac1{4\hbar}\frac{d\lagrangian_\energy}{dt} = 0.
\end{align}
This relation exhibits a constant of motion, which demonstrates that the energy-optimal dynamics is that whose energy presents a constant argument of integration $\lagrangian_\energy=\hbar\lagrangian^2=\hbar\Omega^2$, where $\Omega$ is a constant Rabi frequency, i.e
\begin{align}
\label{etadotOmega}
%4\dot\eta^2\Bigl[\bigl(\dot{\widetilde\phi}\bigr)^2+\cos^2\widetilde\phi\Bigr] &= \Omega^2\nonumber\\
2\lvert\dot\eta\rvert\sqrt{\bigl(\dot{\widetilde\phi}\bigr)^2+\cos^2\widetilde\phi} &=\sqrt{\Omega_P^2+\Omega_S^2}= \Omega=const.,
\end{align}
thus, the partial pulse area is given by:
\begin{equation}
\label{areaEqOmT}
\widetilde\area(t) = \int_{t_i}^tdt\,\lagrangian = \int_{t_i}^tdt\,\Omega = \Omega\,(t-t_i).
\end{equation}
%turns out to be equivalent to that of a field of constant amplitude $\Omega$.
Furthermore, for a given Rabi frequency $\Omega$, the optimal time is then
\begin{equation}
T_\textrm{opt}=\widetilde\area(t_f)/\Omega=\area/\Omega.
\end{equation}
Using $\dot\phi=\dot\eta\dot{\widetilde\phi}$, $\ddot\phi=\dot\eta^2\ddot{\widetilde\phi}+\ddot\eta\dot{\widetilde\phi}$, \eqref{etadotOmega}, and \eqref{EL1_}, \eqref{EL2_} reads
\begin{align}
\label{EL12_}
0 &= -\sgn\dot\eta\frac{\ddot{\widetilde\phi}+\Bigl[2(\dot{\widetilde\phi})^2+\cos^2\widetilde\phi\Bigr]\tan\widetilde\phi}{\Bigl[\bigl(\dot{\widetilde\phi}\bigr)^2+\cos^2\widetilde\phi\Bigr]^{3/2}}\nonumber\\
&\qquad{}+\frac{1}{2\hbar\Omega}\biggl(\mu_0\sec\widetilde\phi+\mu_1\sin\eta-\mu_2\cos\eta\biggr),
\end{align}
where $\mu_0$ has been redefined as $\mu_0/2\rightarrow\mu_0$.
The latter is exactly \eqref{EulerLagrangePhiOfEta} for $\mu_j=2\lambda_j\hbar\Omega$, i.e. gives the same trajectory as for the pulse-area optimization.

Equation \eqref{etadotOmega} can be rewritten as
\begin{align}
\label{etadotOmega_}
\widetilde\area\bigl[\eta(t)\bigr] &= 2\int_{\eta_i}^{\eta(t)}d\eta\,\sgn\dot\eta\sqrt{\bigl(\dot{\widetilde\phi}\bigr)^2+\cos^2\widetilde\phi},\nonumber\\
	&= \Omega\,(t-t_i),
\end{align}
for a trajectory $\widetilde{\phi}(\eta)$, where the left-hand side is the partial generalized pulse area, see Eq.~\eqref{GeneralPulseAreaIntEta}, i.e.~$\Omega=\area/(t_f-t_i)$.
%The form of the right-hand side implies that the generalized amplitude is, in fact, constant, i.e.
%\begin{align}
%&\int_{t_i}^tdt'\,\sqrt{P(t')^2+S(t')^2}\nonumber\\
%&\qquad{}=\int_{t_i}^tdt'\,\sqrt{\Omega^2\sin^2\vartheta(t')+\Omega^2\cos^2\vartheta(t')}.
%\end{align}
As a consequence, the energy-optimal dynamics for the trajectory \eqref{EulerLagrangePhiOfEta} is also its time-optimal solution for a given constant Rabi frequency $\Omega=\Omega_0$, i.e $(t_f-t_i)\rvert_\text{opt}=\area/\Omega_0$.

Enforcing \eqref{etadotOmega} and \eqref{EulerLagrangePhiOfEta} guarantees the resultant dynamics to be globally area-, energy-, and time-optimal.
\section{Discussion}
\label{sec_dicussion}
\begin{figure}
\centering
\includegraphics[scale=1]{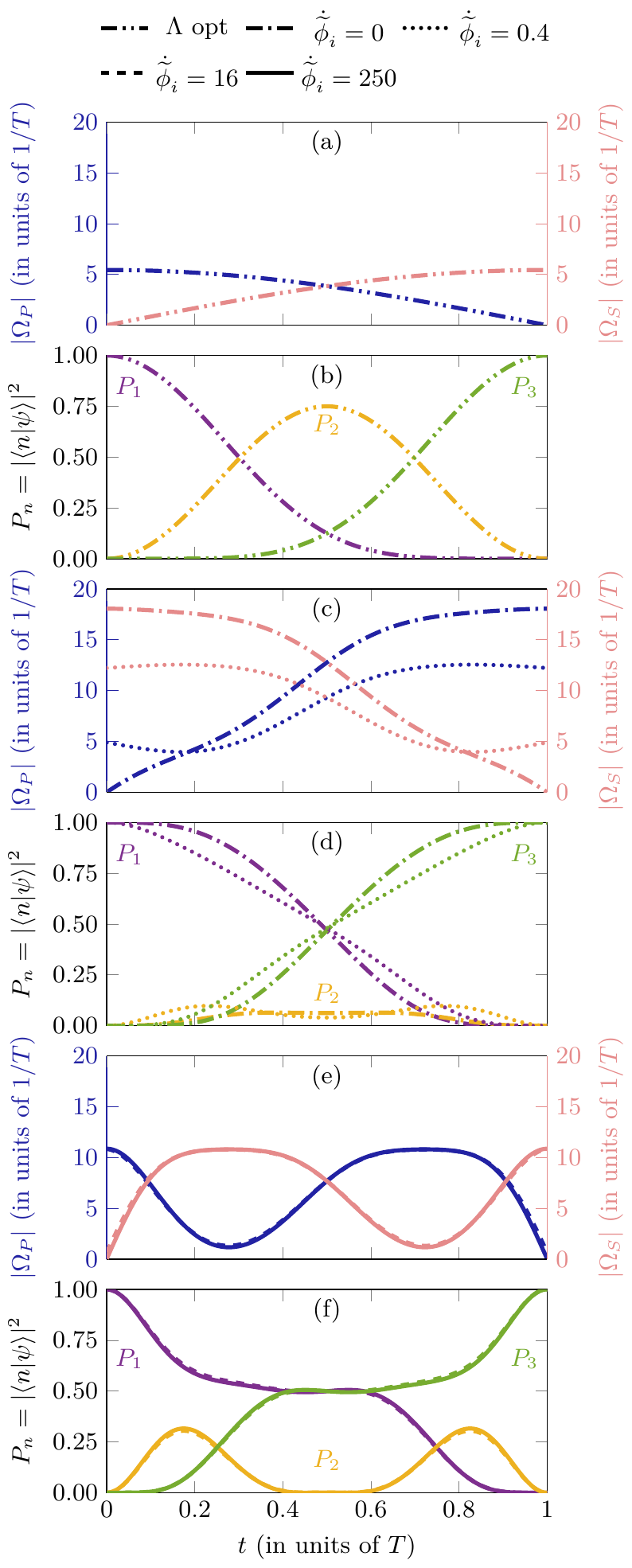}
\caption{Fields and populations vs time [$(a)$ and $(b)$] for unconstrained optimum $\Lambda$ transfer and [$(c)$--$(f)$] for selected robust optima.
Numerical parameters details are summarized in Table \ref{tab_loss_vs_area}.
The line style in the legend applies to all plots irrespective of the line color.
%We notice that the difference between the dynamics for $\dot{\widetilde\phi}_i=16$ and $\dot{\widetilde\phi}_i=250$ is almost unnoticeable at the scale of the figure.
The dynamics for $\dot{\widetilde\phi}_i=16$ and $\dot{\widetilde\phi}_i=250$ are almost indistinguishable at the scale of the figure.}\label{fig_FieldsAndPopulationsVsTime}
\end{figure}

Fields and populations dynamics are shown in Fig.~\ref{fig_FieldsAndPopulationsVsTime}.
The double-dot--dashed lines represent the optimal solution for the problem of population transfer without constraints of robustness, the fields are intuitively-ordered cosine-sine pulses \eqref{cos-sin}.
For the unconstrained optimization the transient population of the excited state is much larger than for the robust extrema.
It can be observed that, while the unconstrained solution transfers most of the population from $\lvert1\rangle$ to $\lvert2\rangle$ and $\lvert3\rangle$, with predominance of the excited state, in the first half of the process, the optimal robust solution transfers only about 40\% of the population from the initial state to a superposition of the others (with much lower transient population on the excited state), proceeds to deplete the excited state into the target state, and only then executes the last part of the transfer like the first part.
In this manner, while the unconstrained optimal transiently populates the excited state along the transfer, the robust optimal solution only does this in two temporally separated stages of 40\% of the process duration and maintains it depopulated 20\% of the time.
At the middle of the process we obtain maximal superposition of the ground states and no population on the excited state, showing that the robust optimal full transfer appears as a two consecutive unconstrained (non-robust) half transfers (as already noticed in the preceding section).

The pulses corresponding to the extrema of the robust optimization go from a counter-intuitively--ordered pair of fields, to what could be described as a train of nonvanishing pulse pairs with intuitive--counter\nobreakhyphen intuitive--intuitive orderings.
%The time optimization for a fixed loss, calculated elsewhere with no robustness constraints, produces pulses that follow the same general idea of ordering.
One can remark that composite-pulses STIRAP exhibits precisely the opposite ordering \cite{Torosov2013}.
From the intermediate extremum it can be seen that the optimum is achieved by raising the bounded boundary of the field towards its maximum, $\Omega$, and lowering the other boundary towards 0.
All field pairs are complementary to each other and exhibit, as a whole, symmetry around half of the duration.

%A point that should be remarked is that by displaying an intuitive ordering at initial and final time intervals with nonzero pump and Stokes, respectively, the fields must be made zero ``instantly'' for $t=t_i-dt$ and $t=t_f+dt$, otherwise the beginning of two-level Rabi oscillations would reduce the fidelity of the transfer.
 
\begin{figure}
\centering
\includegraphics[scale=1]{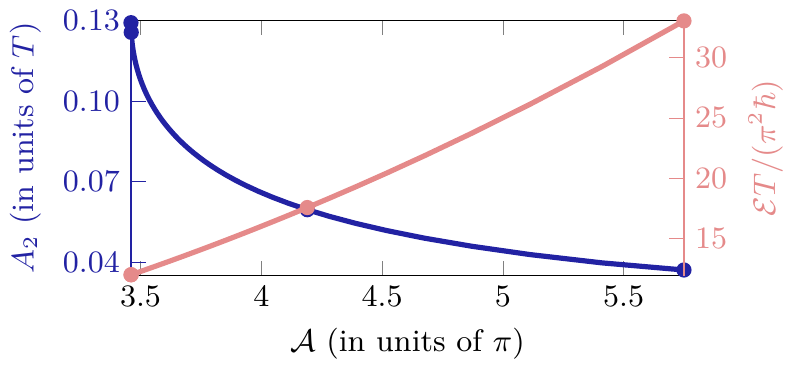}
\caption{Loss-proportional term $A_2$ vs generalized pulse area for optimal robust family of solutions.
To compare with $A_2=(3/8)T=0.3750T$ at $\area=\sqrt{3}\pi\approx1.7321\pi$ and $\energy=3\pi^2\hbar/T$ for the optimal $\Lambda$ transfer with no constraints of robustness.
Numerical details have been gathered in Table \ref{tab_loss_vs_area}.}\label{fig_A2vsArea}
\end{figure}

Now that we have the time-dependence of the angles, we can calculate the loss term $A_2$, we show its dependence with the generalized pulse area in Fig.~\ref{fig_A2vsArea}.
The minimum energy for the produced losses, calculated elsewhere with no robustness constraints, is between $1.92\pi\lesssim\area\lesssim2\pi$ or $5.43\lesssim\energy T/(\pi^2\hbar)\lesssim27$ with mostly overlapping pulses.

The fact that, in addition to the robust optimal solution, we have obtained a family of solutions which perform the desired task robustly and for fairly low areas (compared to adiabatic procedures with $\area>10\pi$) suggests the following practical strategy:
These solutions could be used as options that become more or less preferable depending on the constraints of the implementation, e.g., when a $A_2=0.1291T$ is an acceptable cost, the actual optimal robust solution may be used; however, when that is too high to be acceptable, a larger-area optimum may be chosen, effectively lowering the associated loss term as low as $A_2=0.0371T$.
Recalling Fig.~\ref{fig_robustnessprofile}, we can highlight that for roughly the same robustness profile we can choose among the extrema solutions, varying in area and loss parameter, according with the physical limitations of a particular practical implementation and the loss that can be considered acceptable.

\begin{table}
\caption{Identifying parameters of the highlighted extrema.}\label{tab_loss_vs_area}
\begin{center}
\begin{tabular}{c|ccccc}
             & $\Lambda_{\textrm{opt}}$ & $\dot{\widetilde\phi}_i=0$ & $\dot{\widetilde\phi}_i=0.4$ & $\dot{\widetilde\phi}_i=16$ & $\dot{\widetilde\phi}_i=250$ \\ \hline
$A_2/T$      & 0.3750        &  0.0371                    &  0.0596                      &  0.1256                     &  0.1291   \\ 
$\area/\pi$  & 1.7321        &  5.7498                    &  4.1904                      &  3.4615                     &  3.4603   \\ 
$\dfrac{\energy T}{\pi^2\hbar}$ & 3     &  33.0599      & 17.5597                      & 11.9819                     & 11.9739   \\ 
$\Delta\epsilon_{\textrm{uhf}}$   & 0.4\% &   6.4\%       &  5.6\%                       &  5.1\%                      &  5.1\%    \\
$\lambda_0$  &              &  0.1930790914              &  0.14013                     & -0.52403                    & -0.56596  \\ 
$\lambda_1$  & 0             & -0.0838224029              & -0.04747                     &  0.86793                    &  0.93853  \\ 
$\lambda_2$  & 0             &  0.0102504990              &  0.22984                     &  1.05836                    &  1.08283  \\
$\eta_f/\pi$ & 			 &  2.9225                    &  2.1297                      &  1.5627                     &  1.5454   
\end{tabular}
\end{center}
\end{table}

The corresponding loss parameters and pulse areas, together with pulse energy and the values of the Lagrange multipliers, for the selected extrema in Figs.~\ref{fig_FieldsAndPopulationsVsTime} and \ref{fig_A2vsArea} are presented in Table \ref{tab_loss_vs_area}.
%The optimization of the $\Lambda$ transfer unconstrained by robustness conditions, though an analytical solution obtained by other means, must have a numerical equivalent in our approach.
%In this case, the Lagrange multipliers controlling the robustness are suppressed (made zero) and a value of $\lambda_0$ must reproduce accurately the $\sqrt{3}\pi$--cosine-sine---pulses when $\dot{\widetilde\phi}_i\to\infty$; this is why a question mark is left in the corresponding cell of the table.
The shown precision of the Lagrange multipliers is the minimum necessary to produce the selected results with their displayed precision (while guaranteeing $\widetilde O_2<10^{-4}$).
It can be noted that the largest area extrema requires the most precision on the multipliers, this is only due to the rapid dependence of the areas for small $\dot{\widetilde\phi}_i$'s.

The existence of a continuous family of optima of robust solutions, controlled mathematically by the quantity $\dot{\widetilde\phi}_i$, but interpreted physically as the consequent loss, as extracted from Figs.~\ref{fig_lambdaJandAreavsDphii} and \ref{fig_A2vsArea}, is a remarkable result of the applied method of inverse optimization with robustness as constraints.
For the chosen family of robust optimal solutions, we have control of the loss parameter $A_2$ via the pulse area (equivalently, energy) and, like for adiabatic protocols, the relation is inversely proportional: lower loss requires higher energies, although, unlike the adiabatic behavior, there are lower and upper bounds to them.

We can use the relation \eqref{loss} to estimate the upper bound on the time duration or lower bound on the pulse amplitude of the robust optimal pulses (absolute optimum and largest-area extrema) with respect to the dissipation parameter $\Gamma$ (inverse of relaxation time) considering an admissible loss below the ultrahigh-fidelity benchmark $P_{\mathrm{loss}}\lesssim10^{-4}$, i.e.~$T\lesssim7.7\times10^{-4}\Gamma^{-1}$ and (consequently) $\Omega\gtrsim1.4\times10^4\Gamma$, for $\area=3.4603\pi$, else $T\lesssim27.0\times10^{-4}\Gamma^{-1}$, $\Omega\gtrsim 0.7\times10^4\Gamma$, for $\area=5.7498\pi$.
\section{Conclusions}
We have demonstrated a method to obtain robust quantum transfers while optimizing area, energy, and time.
We have presented, for the first time, the optimal resonant $\Lambda$ transfer with robustness up to the third order in terms of field inhomogeneities.
The resultant pulse shapes are smooth and very energy economical, far below requirements of STIRAP, while also exhibiting a robust behavior comparable to robust STIREP and STIRAP \cite{Laforgue2019} but with $2\pi\sim3\pi$ lower areas.
Losses remain below the ultrahigh-fidelity benchmark of $10^{-4}$ for pulse durations in the order of $\Gamma^{-1}\times10^{-4}$, e.g., for the relaxation time of the excited state $1\mathrm{D}_2$ of a praseodymium ion in a $\mathrm{Pr}^{3+}$:$\mathrm{Y}_2\mathrm{SiO}_5$ crystal, which is $\Gamma^{-1}=T_1\approx164\mu\mathrm{s}$ \cite{Bruns2018,Schraft2013}, the largest pulse duration and smallest amplitude necessary to avoid dissipation losses with the global optimal robust fields are $0.13\mu\mathrm{s}$ and $85.4\mathrm{MHz}$, respectively.

The method could be extended to the search of robustness with respect to detuning and for field inhomogeneities unequal between the fields, and to consider higher orders of robustness.
%The method does not provide analytical solutions, but numerical trajectories that can be made as accurate as required.
%Fits can be used.
%Higher orders robustness ? Robustness with respect to detuning ?

The protocol to produce the pulses can be summarized as follows, focusing on the representative values in Table \ref{tab_loss_vs_area}: 
\begin{enumerate}
\item The values of $\lambda_0$, $\lambda_1$, $\lambda_2$, $\eta_f$, and the pulse area $\mathcal{A}$ for the chosen representative values of $\dot{\widetilde\phi}_i$ allow the user to choose between the members of the robust candidates for optimal pulses, for a given admissible loss of the problem.
\item Once $\dot{\widetilde\phi}_i$ is chosen, taking the other properties summarized in Table \ref{tab_loss_vs_area} as deciding factors, the values of the $\lambda_n$'s, $\eta_f$, and pulse area are also taken and $\widetilde\phi(\eta)$, $\dot{\widetilde\phi}(\eta)$, and $\widetilde{\theta}(\eta)$ are calculated by 
\begin{enumerate}
\item introducing the $\lambda_n$'s in \eqref{numerical_implementation}, which can then be solved using a standard method for solving ordinary differential equations,
\item the system is solved for the chosen $\dot{\widetilde\phi}_i$, with $\widetilde\phi_i=\eta_i=0$ as additional initial conditions, ``integrating'' up to $\eta_f$.
\end{enumerate}	
\item $\dot\eta(t)$ is then obtained, in units of $1/T$, from \eqref{etadotOmega} choosing the maximum pulse amplitude to be $\Omega=\mathcal{A}/T$.
\item $\widetilde\phi(\eta)$, $\dot{\widetilde\phi}(\eta)$, $\widetilde{\theta}(\eta)$, and $\dot\eta(t)$ are finally introduced in \eqref{dynscaledfields} to obtain the pump and Stokes pulses, $\Omega_P(t)$ and $\Omega_S(t)$, in units of $1/T$.
\end{enumerate}
Other values than the ones of Table \ref{tab_loss_vs_area} can be used with Fig.~\refsubcap{fig_lambdaJandAreavsDphii}{(a)} and $\eta_f$ given by \eqref{cosetaf} where $\eta_f\in[0,2\pi)$ for $\dot{\widetilde\phi}_i>0.5798$ and $\eta_f\in[2\pi,4\pi)$ for $0\leq\dot{\widetilde\phi}_i\leq 0.5798$.
\section{Acknowledgments}
This work was supported by the ``Investissements d'Avenir'' program, EUR-EIPHI Graduate School (17-EURE-0002). X.L.~and S.G.~also acknowledge support from the European Union's Horizon 2020 research and innovation program under the Marie Sklodowska-Curie grant agreement No.~765075 (LIMQUET).
\appendix
\section{Property of iterated integrals}
\label{AppIntegral}
Double iterated integrals may be simplified as follows 
\begin{align}
\label{Propinteg2}
&\int_{t_i}^{t_f}dt\!\!\int_{t_i}^tdt'\,[a(t)b(t')
+a(t')b(t)]\nonumber\\
&\qquad=\int_{t_i}^{t_f}dt\,a(t)\int_{t_i}^{t_f}dt\,b(t),
\end{align}
from 
\begin{equation}
\int_{t_i}^{t_f}du(t)\,v(t) + \int_{t_i}^{t_f}dv(t)\,u(t) = [u(t)v(t)]_{t_i}^{t_f},
\end{equation}
where
\begin{subequations}
\begin{alignat}{2}
u(t)&=\int_{t_i}^tdt'\,a(t'),& \quad du &= dt\,a(t),\\
v(t)&=\int_{t_i}^tdt'\,b(t'),& dv &= dt\,b(t).
\end{alignat}
\end{subequations}
For the triple iterated integrals, we have
\begin{align}
&\int_{t_i}^{t_f}dt\,a(t)\!\!\int_{t_i}^{t}dt'\,b(t')\!\!\int_{t_i}^{t'}dt''\,c(t'')\nonumber\\
&\qquad = \int_{t_i}^{t_f}du(t)\!\!\int_{t_i}^{t}dt'\,b(t')w(t')\nonumber\\
&\qquad = \int_{t_i}^{t_f}du(t)\,x(t)=[u(t)x(t)]_{t_i}^{t_f}-\int_{t_i}^{t_f}dx(t)\,u(t)\nonumber\\
&\qquad = \int_{t_i}^{t_f}dt'\,a(t')\int_{t_i}^{t_f}dt'\,b(t')w(t')\nonumber\\
&\qquad \qquad{}-\int_{t_i}^{t_f}dt\,u(t)b(t)w(t),
\end{align}
with
\begin{subequations}
\begin{alignat}{2}
w(t)&=\int_{t_i}^tdt'\,c(t'),&\quad dw &= dt\,c(t),\\
x(t)&=\int_{t_i}^tdt'\,b(t')w(t'),& dx &= dt\,b(t)w(t).
\end{alignat}
\end{subequations}
\section{Numerical implementation}
\label{AppNumImpl}
For a given set $\bigl(\dot{\widetilde\phi}{}^\pm_i,\lambda_0,\lambda_1,\lambda_2\bigr)$, the differential equation \eqref{phitrajectory} is solved numerically from $\eta=\eta_i=0$ to $\eta=3\pi$ (a large enough bound for the low pulse areas we are looking for). 
%$\eta=\eta_f$ satisfying \eqref{cosetaf}. 
A search in this parameter space is then launched in order to find a certain set such that $\widetilde\phi=0$ is satisfied at some point $\eta>0$ which we then denote with the coordinate $(\eta_f,\phi_f)$, while the constraints \eqref{ConstraintsPhiOfEta} are also fulfilled.%, $\dot{\widetilde\phi}^\pm_m\to0$, and $\dot{\widetilde\phi}^\pm_f=-\dot{\widetilde\phi}^\pm_i$ (shooting method).

The implementation for the numerical resolution, via a solver using the Runge-Kutta method, is then
\begin{subequations}
\label{numerical_implementation}
\begin{alignat}{2}
\dot y_1 &= \dot{\widetilde\phi}{}^\pm &&= y_2,\\
\dot y_2 &= \ddot{\widetilde\phi}{}^\pm &&= -\bigl(2y_2^2+\cos^2\widetilde\phi^\pm\bigr)\tan\widetilde\phi^\pm\nonumber\\
&&&\pm\bigl(\lambda_0\sec\widetilde\phi^\pm+\lambda_1\sin\eta-\lambda_2\cos\eta\bigr)\nonumber\\
&&&\qquad{}\times\bigl(y_2^2+\cos^2\widetilde\phi^\pm\bigr)^{3/2}=0.
\end{alignat}
\end{subequations}
The numerical solutions show that the derived optimal family of trajectories is actually symmetric, implying that the assumptions in Appendix \ref{Appsymmtraj} are valid and that Eqs.~\eqref{cosetaf} are satisfied.
\section{Symmetric trajectory}
\label{Appsymmtraj}
In this Appendix we consider a symmetric solution via the standard forward evolution of \eqref{phitrajectory} and its backward counterpart, i.e.~we consider the trajectory evolving from $\eta=0$ to $\eta=\eta_f$ and its reversal starting from the end point of the trajectory and ending at the starting point.
The backward-propagating equation is obtained making $\widetilde\phi^\pm(\eta)\xrightarrow{\eta=\eta_f-u}\widehat\phi^\pm(u)$, where $\widehat\phi^\pm(u)$ is the backward-propagating trajectory and $u$ is the counterpart of $\eta$ (i.e.~identical to $\eta$ but time-reversed).
Consequently, $\dot{\widehat\phi}{}^\pm=-\dot{\widetilde\phi}{}^\pm$ and $\ddot{\widehat\phi}{}^\pm=\ddot{\widetilde\phi}{}^\pm$.
Then,
\begin{align}
\label{backwardPhiOfEta}
&\mp\frac{\ddot{\widehat\phi}{}^\pm+\Bigl[2\bigl(\dot{\widehat\phi}{}^\pm\bigr)^2+\cos^2\widehat\phi^\pm\Bigr]\tan\widehat\phi^\pm}{\Bigl[\bigl(\dot{\widehat\phi}{}^\pm\bigr)^2+\cos^2\widehat\phi^\pm\Bigr]^{3/2}}+\lambda_0\sec\widehat\phi^\pm\nonumber\\
&\qquad{}+\lambda_1\sin(\eta_f-u)-\lambda_2\cos(\eta_f-u)=0.
\end{align}
The symmetric solution is implemented demanding symmetry about the axis $\eta(t=t_i+T/2)=\eta_m=\eta_f/2$, where $T$ is the pulse duration, implying $\widehat\phi^\pm(u)=\widetilde\phi^\pm(\eta)$ (i.e.~the equality of the counterpropagating trajectories), which leads to:
%\begin{align}
%\lambda_1\sin\eta-\lambda_2\cos\eta &= \lambda_1\sin(\eta_f-\eta)-\lambda_2\cos(\eta_f-\eta),%\nonumber\\
%%	&= -(\lambda_1\cos\eta_f+\lambda_2\sin\eta_f)\sin\eta\nonumber\\
%%	&\qquad{}-(\lambda_2\cos\eta_f-\lambda_1\sin\eta_f)\cos\eta,
%\end{align} 
%i.e.
%\begin{subequations}
%\label{cos_sin_lam}
%\begin{align}
% \lambda_1 &= -\lambda_1\cos\eta_f -\lambda_2\sin\eta_f,\\
%-\lambda_2 &= -\lambda_2\cos\eta_f +\lambda_1\sin\eta_f,
%\end{align} 
%\end{subequations}
%i.e.
\begin{subequations}
\label{lam_cos_sin}
\begin{align}
0 &= \lambda_1(1+\cos\eta_f)+\lambda_2\sin\eta_f,\\
0 &= \lambda_1\sin\eta_f+\lambda_2(1-\cos\eta_f).
\end{align} 
\end{subequations}
From Eqs.~\eqref{lam_cos_sin}, we have $\lambda_{1,2}\ne0$ when the determinant of the right-hand side matrix form is zero, i.e. $(1+\cos\eta_f )(1-\cos\eta_f)-\sin^2\eta_f=0$, which is always satisfied. 
We can alternatively express the cosine and sine as:
\begin{align}
\label{cosetaf}
\cos\eta_f=\frac{\lambda_2^2-\lambda^2_1}{\lambda_1^2 + \lambda^2_2},\quad \sin\eta_f=-\frac{2\lambda_1\lambda_2}{\lambda_1^2 + \lambda^2_2}.
\end{align}
%From \eqref{cosetaf} we get
%\begin{subequations}
%\begin{align}
%\cos\eta_f &= \cos2\eta_m=2\cos^2\eta_m-1=1-2\sin^2\eta_m,\\
%\sin\eta_f &= \sin2\eta_m=2\sin\eta_m\cos\eta_m,
%\end{align}
%\end{subequations}
%i.e.
%\begin{subequations}
%\begin{align}
%\frac{\sin^2\eta_f}{2\cos^2\eta_m} &= 2\sin^2\eta_m = 1-\cos\eta_f \nonumber\\
%	\Longrightarrow\cos^2\eta_m&=\frac{\sin^2\eta_f}{2(1-\cos\eta_f)}=\frac{\lambda_2^2}{\lambda_1^2+\lambda_2^2},\\
%\frac{\sin^2\eta_f}{2\sin^2\eta_m} &= 2\cos^2\eta_m = 1+\cos\eta_f\nonumber\\
%	\Longrightarrow\sin^2\eta_m&=\frac{\sin^2\eta_f}{2(1+\cos\eta_f)}=\frac{\lambda_1^2}{\lambda_1^2+\lambda_2^2},
%\end{align}
%\end{subequations}
%which results in
%\begin{subequations}
%\begin{align}
%\label{cosEtam_sinEtam}
%\lvert\sin\eta_m\rvert &= \frac{\lvert\lambda_1\rvert}{\sqrt{\lambda_1^2+\lambda_2^2}},\\
%\lvert\cos\eta_m\rvert &= \frac{\lvert\lambda_2\rvert}{\sqrt{\lambda_1^2+\lambda_2^2}}.
%\end{align}
%\end{subequations}
Equation \eqref{cosetaf} gives $\eta_f$ with modulus $2\pi$, hence $\eta_f$ is known once we know in which interval $\eta_{fn}\in\bigl[2n\pi,2(n+1)\pi\bigr)$, with $n=0,\pm1,\pm2,\dots$ (where the sign is fixed by the sign of $\dot\eta$), it is located.
%Moreover, the signs of the sine and cosine of $\eta_m$ are given by the interval where $\eta_f$ is defined to be, i.e.
%\begin{subequations}
%\begin{align}
%\sgn(\sin\eta_m) &= (-1)^n,\\
%\sgn(\cos\eta_m) &= (-1)^{n+1}\sgn(\lambda_1\lambda_2).
%\end{align}
%\end{subequations}
For the obtained cosine and sine in Eq.~\eqref{cosetaf}, the backward\nobreakhyphen propagating equation \eqref{backwardPhiOfEta} becomes identical to the forward\nobreakhyphen propagating one \eqref{phitrajectory} (albeit in terms of $u$).

We may refer to the controls \eqref{fields-Eulerangles} to note that the pulses are mirror images of each other, $\Omega_P(t)=\Omega_S(t_f-t+t_i)$, when we apply the time-reversal symmetry:
\begin{equation}
\theta\to\frac{\pi}{2}-\widehat\theta,\quad\phi\to\widehat\phi,\quad\dot\phi\to-\dot{\widehat\phi},\quad\dot\eta\to-\dot{\widehat\eta}.
\end{equation}
\section{Dynamics of the angles}
\label{AppDynAngles}
\begin{figure}
\centering
\includegraphics[scale=1]{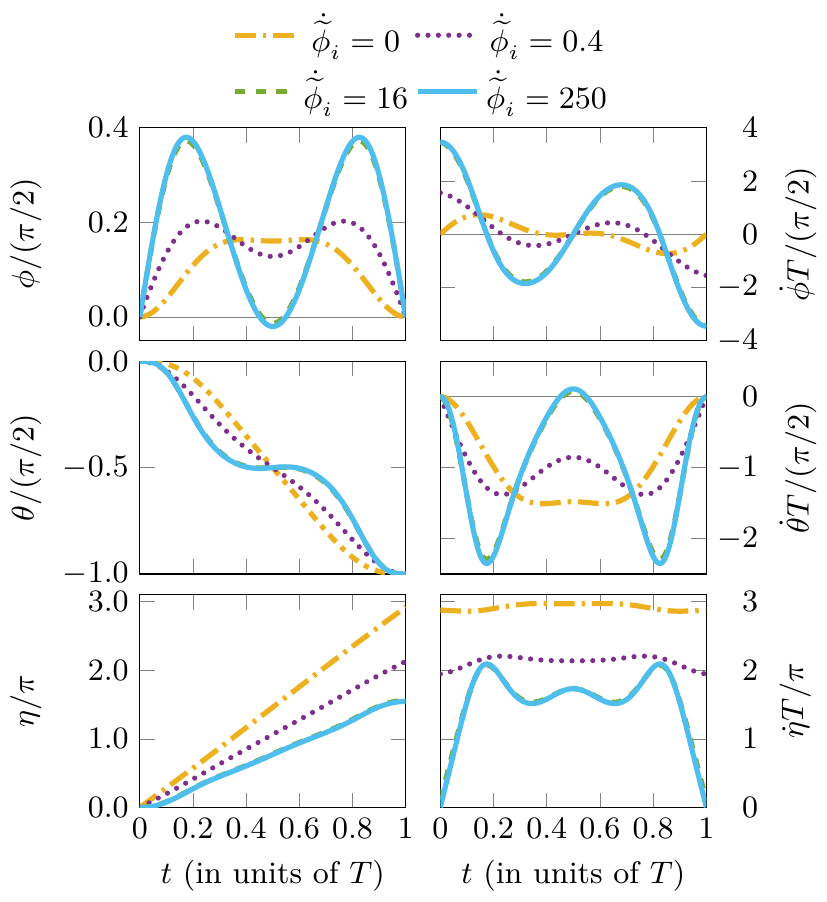}
\caption{Dynamics of the angular parametrization $(\phi,\eta,\theta)$ for the selected optima from Fig.~\ref{fig_lambdaJandAreavsDphii}.
Numerical parameters are summarized on Table \ref{tab_loss_vs_area}.
Thin gray lines mark the zero.
The process duration $T$, equal to the pulse duration, is taken to be common among the extrema by letting the pulse amplitudes be given by $\Omega=\area/T$ for each corresponding area.}\label{fig_PhiThetaAndEtaOfTime}
\end{figure}

The dynamics of the angles and their derivative is presented in Fig.~\ref{fig_PhiThetaAndEtaOfTime}.
The description of the trajectories $\widetilde\phi(\eta)$ and $\widetilde\theta(\eta)$ can be recalled to discuss their corresponding time-dependent evolutions, although it is worth noting that the process duration $T$ for all solutions was made equal only after proper choice of the generalized pulse amplitude $\Omega$.
Unlike for the geometrical trajectory, the time-dependent functions are all finite.
The evolution of $\eta$ is almost a straight line of slope $3/T$ for the maximum-area extrema, however the optimum is obtained when $\eta$ approaches a slightly oscillating line with zero derivative at the boundaries.
%\label{phiisaveven}While $\phi$ is an even function with nonzero derivative at the boundaries, $\dot\eta$ and $\dot\theta$ are even functions null-valued at the boundaries.
All the presented dynamics display a certain parity with respect to $t=T/2$: all functions are mirrored or anti-mirrored (sign-changed) around that point, except $\theta$ and $\eta$. $\theta$ and $\eta$ are odd functions only when also regarded about their value when evaluated at that point, i.e., the functions $f(t)=\theta(t-T/2)-\theta(T/2)$ and $g(t)=\eta(t-T/2)-\eta(T/2)$ are odd. The only function that is not null-valued at neither of its boundaries is $\dot\phi$, giving the pump and Stokes fields their respective nonzero boundary.
%\bibliography{OptRobSTIREP}

\begin{thebibliography}{31}%
\makeatletter
\providecommand \@ifxundefined [1]{%
 \@ifx{#1\undefined}
}%
\providecommand \@ifnum [1]{%
 \ifnum #1\expandafter \@firstoftwo
 \else \expandafter \@secondoftwo
 \fi
}%
\providecommand \@ifx [1]{%
 \ifx #1\expandafter \@firstoftwo
 \else \expandafter \@secondoftwo
 \fi
}%
\providecommand \natexlab [1]{#1}%
\providecommand \enquote  [1]{``#1''}%
\providecommand \bibnamefont  [1]{#1}%
\providecommand \bibfnamefont [1]{#1}%
\providecommand \citenamefont [1]{#1}%
\providecommand \href@noop [0]{\@secondoftwo}%
\providecommand \href [0]{\begingroup \@sanitize@url \@href}%
\providecommand \@href[1]{\@@startlink{#1}\@@href}%
\providecommand \@@href[1]{\endgroup#1\@@endlink}%
\providecommand \@sanitize@url [0]{\catcode `\\12\catcode `\$12\catcode
  `\&12\catcode `\#12\catcode `\^12\catcode `\_12\catcode `\%12\relax}%
\providecommand \@@startlink[1]{}%
\providecommand \@@endlink[0]{}%
\providecommand \url  [0]{\begingroup\@sanitize@url \@url }%
\providecommand \@url [1]{\endgroup\@href {#1}{\urlprefix }}%
\providecommand \urlprefix  [0]{URL }%
\providecommand \Eprint [0]{\href }%
\providecommand \doibase [0]{https://doi.org/}%
\providecommand \selectlanguage [0]{\@gobble}%
\providecommand \bibinfo  [0]{\@secondoftwo}%
\providecommand \bibfield  [0]{\@secondoftwo}%
\providecommand \translation [1]{[#1]}%
\providecommand \BibitemOpen [0]{}%
\providecommand \bibitemStop [0]{}%
\providecommand \bibitemNoStop [0]{.\EOS\space}%
\providecommand \EOS [0]{\spacefactor3000\relax}%
\providecommand \BibitemShut  [1]{\csname bibitem#1\endcsname}%
\let\auto@bib@innerbib\@empty
%</preamble>
\bibitem [{\citenamefont {Gaubatz}\ \emph {et~al.}(1990)\citenamefont
  {Gaubatz}, \citenamefont {Rudecki}, \citenamefont {Schiemann},\ and\
  \citenamefont {Bergmann}}]{Gaubatz1990}%
  \BibitemOpen
  \bibfield  {author} {\bibinfo {author} {\bibfnamefont {U.}~\bibnamefont
  {Gaubatz}}, \bibinfo {author} {\bibfnamefont {P.}~\bibnamefont {Rudecki}},
  \bibinfo {author} {\bibfnamefont {S.}~\bibnamefont {Schiemann}},\ and\
  \bibinfo {author} {\bibfnamefont {K.}~\bibnamefont {Bergmann}},\ }\bibfield
  {title} {\bibinfo {title} {{Population transfer between molecular vibrational
  levels by stimulated Raman scattering with partially overlapping laser
  fields. A new concept and experimental results}},\ }\href
  {https://doi.org/10.1063/1.458514} {\bibfield  {journal} {\bibinfo  {journal}
  {J.~Chem.~Phys.~}\ }\textbf {\bibinfo {volume} {92}},\ \bibinfo {pages}
  {5363} (\bibinfo {year} {1990})}\BibitemShut {NoStop}%
\bibitem [{\citenamefont {Kuhn}\ \emph {et~al.}(2002)\citenamefont {Kuhn},
  \citenamefont {Hennrich},\ and\ \citenamefont {Rempe}}]{Kuhn2002}%
  \BibitemOpen
  \bibfield  {author} {\bibinfo {author} {\bibfnamefont {A.}~\bibnamefont
  {Kuhn}}, \bibinfo {author} {\bibfnamefont {M.}~\bibnamefont {Hennrich}},\
  and\ \bibinfo {author} {\bibfnamefont {G.}~\bibnamefont {Rempe}},\ }\bibfield
   {title} {\bibinfo {title} {{Deterministic Single-Photon Source for
  Distributed Quantum Networking}},\ }\href
  {https://doi.org/10.1103/PhysRevLett.89.067901} {\bibfield  {journal}
  {\bibinfo  {journal} {Phys. Rev. Lett.}\ }\textbf {\bibinfo {volume} {89}},\
  \bibinfo {pages} {067901} (\bibinfo {year} {2002})}\BibitemShut {NoStop}%
\bibitem [{\citenamefont {S{\o}rensen}\ \emph {et~al.}(2006)\citenamefont
  {S{\o}rensen}, \citenamefont {M{\o}ller}, \citenamefont {Iversen},
  \citenamefont {Thomsen}, \citenamefont {Jensen}, \citenamefont {Staanum},
  \citenamefont {Voigt},\ and\ \citenamefont {Drewsen}}]{Sorensen2006}%
  \BibitemOpen
  \bibfield  {author} {\bibinfo {author} {\bibfnamefont {J.~L.}\ \bibnamefont
  {S{\o}rensen}}, \bibinfo {author} {\bibfnamefont {D.}~\bibnamefont
  {M{\o}ller}}, \bibinfo {author} {\bibfnamefont {T.}~\bibnamefont {Iversen}},
  \bibinfo {author} {\bibfnamefont {J.~B.}\ \bibnamefont {Thomsen}}, \bibinfo
  {author} {\bibfnamefont {F.}~\bibnamefont {Jensen}}, \bibinfo {author}
  {\bibfnamefont {P.}~\bibnamefont {Staanum}}, \bibinfo {author} {\bibfnamefont
  {D.}~\bibnamefont {Voigt}},\ and\ \bibinfo {author} {\bibfnamefont
  {M.}~\bibnamefont {Drewsen}},\ }\bibfield  {title} {\bibinfo {title}
  {{Efficient coherent internal state transfer in trapped ions using stimulated
  Raman adiabatic passage}},\ }\href
  {https://doi.org/10.1088/1367-2630/8/11/261} {\bibfield  {journal} {\bibinfo
  {journal} {New J. Phys.}\ }\textbf {\bibinfo {volume} {8}},\ \bibinfo {pages}
  {261} (\bibinfo {year} {2006})}\BibitemShut {NoStop}%
\bibitem [{\citenamefont {Kr{\'{a}}l}\ \emph {et~al.}(2007)\citenamefont
  {Kr{\'{a}}l}, \citenamefont {Thanopulos},\ and\ \citenamefont
  {Shapiro}}]{Kral2007}%
  \BibitemOpen
  \bibfield  {author} {\bibinfo {author} {\bibfnamefont {P.}~\bibnamefont
  {Kr{\'{a}}l}}, \bibinfo {author} {\bibfnamefont {I.}~\bibnamefont
  {Thanopulos}},\ and\ \bibinfo {author} {\bibfnamefont {M.}~\bibnamefont
  {Shapiro}},\ }\bibfield  {title} {\bibinfo {title} {{Colloquium: Coherently
  controlled adiabatic passage}},\ }\href
  {https://doi.org/10.1103/RevModPhys.79.53} {\bibfield  {journal} {\bibinfo
  {journal} {Rev. Mod. Phys.}\ }\textbf {\bibinfo {volume} {79}},\ \bibinfo
  {pages} {53} (\bibinfo {year} {2007})}\BibitemShut {NoStop}%
\bibitem [{\citenamefont {Bergmann}\ \emph {et~al.}(2015)\citenamefont
  {Bergmann}, \citenamefont {Vitanov},\ and\ \citenamefont
  {Shore}}]{Bergmann2015}%
  \BibitemOpen
  \bibfield  {author} {\bibinfo {author} {\bibfnamefont {K.}~\bibnamefont
  {Bergmann}}, \bibinfo {author} {\bibfnamefont {N.~V.}\ \bibnamefont
  {Vitanov}},\ and\ \bibinfo {author} {\bibfnamefont {B.~W.}\ \bibnamefont
  {Shore}},\ }\bibfield  {title} {\bibinfo {title} {Perspective: {S}timulated
  {R}aman adiabatic passage: {T}he status after 25 years},\ }\href
  {https://doi.org/10.1063/1.4916903} {\bibfield  {journal} {\bibinfo
  {journal} {J.~Chem.~Phys.}\ }\textbf {\bibinfo {volume} {142}},\ \bibinfo
  {pages} {170901} (\bibinfo {year} {2015})}\BibitemShut {NoStop}%
\bibitem [{\citenamefont {Vitanov}\ \emph {et~al.}(2017)\citenamefont
  {Vitanov}, \citenamefont {Rangelov}, \citenamefont {Shore},\ and\
  \citenamefont {Bergmann}}]{Vitanov2017}%
  \BibitemOpen
  \bibfield  {author} {\bibinfo {author} {\bibfnamefont {N.~V.}\ \bibnamefont
  {Vitanov}}, \bibinfo {author} {\bibfnamefont {A.~A.}\ \bibnamefont
  {Rangelov}}, \bibinfo {author} {\bibfnamefont {B.~W.}\ \bibnamefont
  {Shore}},\ and\ \bibinfo {author} {\bibfnamefont {K.}~\bibnamefont
  {Bergmann}},\ }\bibfield  {title} {\bibinfo {title} {Stimulated {R}aman
  adiabatic passage in physics, chemistry, and beyond},\ }\href
  {https://doi.org/10.1103/RevModPhys.89.015006} {\bibfield  {journal}
  {\bibinfo  {journal} {Rev.~Mod.~Phys.}\ }\textbf {\bibinfo {volume} {89}},\
  \bibinfo {pages} {015006} (\bibinfo {year} {2017})}\BibitemShut {NoStop}%
\bibitem [{\citenamefont {Shore}(2011)}]{Shore2011}%
  \BibitemOpen
  \bibfield  {author} {\bibinfo {author} {\bibfnamefont {B.~W.}\ \bibnamefont
  {Shore}},\ }\href@noop {} {\emph {\bibinfo {title} {Manipulating Quantum
  Structures Using Laser Pulses}}}\ (\bibinfo  {publisher} {Cambridge
  University Press},\ \bibinfo {year} {2011})\BibitemShut {NoStop}%
\bibitem [{\citenamefont {Dridi}\ \emph {et~al.}(2009)\citenamefont {Dridi},
  \citenamefont {Gu{\'{e}}rin}, \citenamefont {Hakobyan}, \citenamefont
  {Jauslin},\ and\ \citenamefont {Eleuch}}]{Dridi2009}%
  \BibitemOpen
  \bibfield  {author} {\bibinfo {author} {\bibfnamefont {G.}~\bibnamefont
  {Dridi}}, \bibinfo {author} {\bibfnamefont {S.}~\bibnamefont {Gu{\'{e}}rin}},
  \bibinfo {author} {\bibfnamefont {V.}~\bibnamefont {Hakobyan}}, \bibinfo
  {author} {\bibfnamefont {H.~R.}\ \bibnamefont {Jauslin}},\ and\ \bibinfo
  {author} {\bibfnamefont {H.}~\bibnamefont {Eleuch}},\ }\bibfield  {title}
  {\bibinfo {title} {Ultrafast stimulated raman parallel adiabatic passage by
  shaped pulses},\ }\href {https://doi.org/10.1103/physreva.80.043408}
  {\bibfield  {journal} {\bibinfo  {journal} {Phys.~Rev.~A}\ }\textbf {\bibinfo
  {volume} {80}},\ \bibinfo {pages} {043408} (\bibinfo {year}
  {2009})}\BibitemShut {NoStop}%
\bibitem [{\citenamefont {Ib{\'a}{\~{n}}ez}\ \emph {et~al.}(2013)\citenamefont
  {Ib{\'a}{\~{n}}ez}, \citenamefont {Chen},\ and\ \citenamefont
  {Muga}}]{Ibanez2013}%
  \BibitemOpen
  \bibfield  {author} {\bibinfo {author} {\bibfnamefont {S.}~\bibnamefont
  {Ib{\'a}{\~{n}}ez}}, \bibinfo {author} {\bibfnamefont {X.}~\bibnamefont
  {Chen}},\ and\ \bibinfo {author} {\bibfnamefont {J.~G.}\ \bibnamefont
  {Muga}},\ }\bibfield  {title} {\bibinfo {title} {Improving shortcuts to
  adiabaticity by iterative interaction pictures},\ }\href
  {https://doi.org/10.1103/physreva.87.043402} {\bibfield  {journal} {\bibinfo
  {journal} {Phys.~Rev.~A}\ }\textbf {\bibinfo {volume} {87}},\ \bibinfo
  {pages} {043402} (\bibinfo {year} {2013})}\BibitemShut {NoStop}%
\bibitem [{\citenamefont {Song}\ \emph {et~al.}(2016)\citenamefont {Song},
  \citenamefont {Ai}, \citenamefont {Qiu},\ and\ \citenamefont
  {Deng}}]{Song2016}%
  \BibitemOpen
  \bibfield  {author} {\bibinfo {author} {\bibfnamefont {X.-K.}\ \bibnamefont
  {Song}}, \bibinfo {author} {\bibfnamefont {Q.}~\bibnamefont {Ai}}, \bibinfo
  {author} {\bibfnamefont {J.}~\bibnamefont {Qiu}},\ and\ \bibinfo {author}
  {\bibfnamefont {F.-G.}\ \bibnamefont {Deng}},\ }\bibfield  {title} {\bibinfo
  {title} {Physically feasible three-level transitionless quantum driving with
  multiple schrödinger dynamics},\ }\href
  {https://doi.org/10.1103/physreva.93.052324} {\bibfield  {journal} {\bibinfo
  {journal} {Phys.~Rev.~A}\ }\textbf {\bibinfo {volume} {93}},\ \bibinfo
  {pages} {052324} (\bibinfo {year} {2016})}\BibitemShut {NoStop}%
\bibitem [{\citenamefont {Huang}\ \emph {et~al.}(2016)\citenamefont {Huang},
  \citenamefont {Chen}, \citenamefont {Wu}, \citenamefont {Song},\ and\
  \citenamefont {Xia}}]{Huang2016}%
  \BibitemOpen
  \bibfield  {author} {\bibinfo {author} {\bibfnamefont {B.-H.}\ \bibnamefont
  {Huang}}, \bibinfo {author} {\bibfnamefont {Y.-H.}\ \bibnamefont {Chen}},
  \bibinfo {author} {\bibfnamefont {Q.-C.}\ \bibnamefont {Wu}}, \bibinfo
  {author} {\bibfnamefont {J.}~\bibnamefont {Song}},\ and\ \bibinfo {author}
  {\bibfnamefont {Y.}~\bibnamefont {Xia}},\ }\bibfield  {title} {\bibinfo
  {title} {Fast generating greenberger{\textendash}horne{\textendash}zeilinger
  state via iterative interaction pictures},\ }\href
  {https://doi.org/10.1088/1612-2011/13/10/105202} {\bibfield  {journal}
  {\bibinfo  {journal} {Laser Phys.~Lett.}\ }\textbf {\bibinfo {volume} {13}},\
  \bibinfo {pages} {105202} (\bibinfo {year} {2016})}\BibitemShut {NoStop}%
\bibitem [{\citenamefont {Li}\ and\ \citenamefont {Chen}(2016)}]{Li2016}%
  \BibitemOpen
  \bibfield  {author} {\bibinfo {author} {\bibfnamefont {Y.-C.}\ \bibnamefont
  {Li}}\ and\ \bibinfo {author} {\bibfnamefont {X.}~\bibnamefont {Chen}},\
  }\bibfield  {title} {\bibinfo {title} {Shortcut to adiabatic population
  transfer in quantum three-level systems: Effective two-level problems and
  feasible counterdiabatic driving},\ }\href
  {https://doi.org/10.1103/physreva.94.063411} {\bibfield  {journal} {\bibinfo
  {journal} {Phys.~Rev.~A}\ }\textbf {\bibinfo {volume} {94}},\ \bibinfo
  {pages} {063411} (\bibinfo {year} {2016})}\BibitemShut {NoStop}%
\bibitem [{\citenamefont {Kang}\ \emph {et~al.}(2016)\citenamefont {Kang},
  \citenamefont {Chen}, \citenamefont {Wu}, \citenamefont {Huang},
  \citenamefont {Song},\ and\ \citenamefont {Xia}}]{Kang2016}%
  \BibitemOpen
  \bibfield  {author} {\bibinfo {author} {\bibfnamefont {Y.-H.}\ \bibnamefont
  {Kang}}, \bibinfo {author} {\bibfnamefont {Y.-H.}\ \bibnamefont {Chen}},
  \bibinfo {author} {\bibfnamefont {Q.-C.}\ \bibnamefont {Wu}}, \bibinfo
  {author} {\bibfnamefont {B.-H.}\ \bibnamefont {Huang}}, \bibinfo {author}
  {\bibfnamefont {J.}~\bibnamefont {Song}},\ and\ \bibinfo {author}
  {\bibfnamefont {Y.}~\bibnamefont {Xia}},\ }\bibfield  {title} {\bibinfo
  {title} {Fast generation of w states of superconducting qubits with multiple
  schrödinger dynamics},\ }\href {https://doi.org/10.1038/srep36737}
  {\bibfield  {journal} {\bibinfo  {journal} {Sci.~Rep.}\ }\textbf {\bibinfo
  {volume} {6}},\ \bibinfo {pages} {36737} (\bibinfo {year}
  {2016})}\BibitemShut {NoStop}%
\bibitem [{\citenamefont {Chen}\ and\ \citenamefont {Muga}(2012)}]{Chen2012}%
  \BibitemOpen
  \bibfield  {author} {\bibinfo {author} {\bibfnamefont {X.}~\bibnamefont
  {Chen}}\ and\ \bibinfo {author} {\bibfnamefont {J.~G.}\ \bibnamefont
  {Muga}},\ }\bibfield  {title} {\bibinfo {title} {Engineering of fast
  population transfer in three-level systems},\ }\href
  {https://doi.org/10.1103/physreva.86.033405} {\bibfield  {journal} {\bibinfo
  {journal} {Phys.~Rev.~A}\ }\textbf {\bibinfo {volume} {86}},\ \bibinfo
  {pages} {033405} (\bibinfo {year} {2012})}\BibitemShut {NoStop}%
\bibitem [{\citenamefont {Laforgue}\ \emph {et~al.}(2019)\citenamefont
  {Laforgue}, \citenamefont {Chen},\ and\ \citenamefont
  {Gu{\'e}rin}}]{Laforgue2019}%
  \BibitemOpen
  \bibfield  {author} {\bibinfo {author} {\bibfnamefont {X.}~\bibnamefont
  {Laforgue}}, \bibinfo {author} {\bibfnamefont {X.}~\bibnamefont {Chen}},\
  and\ \bibinfo {author} {\bibfnamefont {S.}~\bibnamefont {Gu{\'e}rin}},\
  }\bibfield  {title} {\bibinfo {title} {Robust stimulated {R}aman exact
  passage using shaped pulses},\ }\href
  {https://doi.org/10.1103/physreva.100.023415} {\bibfield  {journal} {\bibinfo
   {journal} {Phys.~Rev.~A}\ }\textbf {\bibinfo {volume} {100}},\ \bibinfo
  {pages} {023415} (\bibinfo {year} {2019})}\BibitemShut {NoStop}%
\bibitem [{\citenamefont {Liu}\ \emph {et~al.}(2019)\citenamefont {Liu},
  \citenamefont {Song}, \citenamefont {Xue}, \citenamefont {Wang},\ and\
  \citenamefont {Yung}}]{Liu2019}%
  \BibitemOpen
  \bibfield  {author} {\bibinfo {author} {\bibfnamefont {B.-J.}\ \bibnamefont
  {Liu}}, \bibinfo {author} {\bibfnamefont {X.-K.}\ \bibnamefont {Song}},
  \bibinfo {author} {\bibfnamefont {Z.-Y.}\ \bibnamefont {Xue}}, \bibinfo
  {author} {\bibfnamefont {X.}~\bibnamefont {Wang}},\ and\ \bibinfo {author}
  {\bibfnamefont {M.-H.}\ \bibnamefont {Yung}},\ }\bibfield  {title} {\bibinfo
  {title} {Plug-and-play approach to nonadiabatic geometric quantum gates},\
  }\bibfield  {journal} {\bibinfo  {journal} {Physical Review Letters}\
  }\textbf {\bibinfo {volume} {123}},\ \href
  {https://doi.org/10.1103/physrevlett.123.100501}
  {10.1103/physrevlett.123.100501} (\bibinfo {year} {2019})\BibitemShut
  {NoStop}%
\bibitem [{\citenamefont {Liu}\ and\ \citenamefont {Yung}(2021)}]{Liu2021}%
  \BibitemOpen
  \bibfield  {author} {\bibinfo {author} {\bibfnamefont {B.-J.}\ \bibnamefont
  {Liu}}\ and\ \bibinfo {author} {\bibfnamefont {M.-H.}\ \bibnamefont {Yung}},\
  }\bibfield  {title} {\bibinfo {title} {Coherent control with user-defined
  passage},\ }\href {https://doi.org/10.1088/2058-9565/abd5ca} {\bibfield
  {journal} {\bibinfo  {journal} {Quantum Science and Technology}\ }\textbf
  {\bibinfo {volume} {6}},\ \bibinfo {pages} {025002} (\bibinfo {year}
  {2021})}\BibitemShut {NoStop}%
\bibitem [{\citenamefont {Dorier}\ \emph {et~al.}(2017)\citenamefont {Dorier},
  \citenamefont {Gevorgyan}, \citenamefont {Ishkhanyan}, \citenamefont {Leroy},
  \citenamefont {Jauslin},\ and\ \citenamefont {Gu{\'{e}}rin}}]{Dorier2017}%
  \BibitemOpen
  \bibfield  {author} {\bibinfo {author} {\bibfnamefont {V.}~\bibnamefont
  {Dorier}}, \bibinfo {author} {\bibfnamefont {M.}~\bibnamefont {Gevorgyan}},
  \bibinfo {author} {\bibfnamefont {A.}~\bibnamefont {Ishkhanyan}}, \bibinfo
  {author} {\bibfnamefont {C.}~\bibnamefont {Leroy}}, \bibinfo {author}
  {\bibfnamefont {H.}~\bibnamefont {Jauslin}},\ and\ \bibinfo {author}
  {\bibfnamefont {S.}~\bibnamefont {Gu{\'{e}}rin}},\ }\bibfield  {title}
  {\bibinfo {title} {Nonlinear stimulated raman exact passage by
  resonance-locked inverse engineering},\ }\href
  {https://doi.org/10.1103/physrevlett.119.243902} {\bibfield  {journal}
  {\bibinfo  {journal} {Phys.~Rev.~Lett.}\ }\textbf {\bibinfo {volume} {119}},\
  \bibinfo {pages} {243902} (\bibinfo {year} {2017})}\BibitemShut {NoStop}%
\bibitem [{\citenamefont {Daems}\ \emph {et~al.}(2013)\citenamefont {Daems},
  \citenamefont {Ruschhaupt}, \citenamefont {Sugny},\ and\ \citenamefont
  {Gu\'erin}}]{Daems2013}%
  \BibitemOpen
  \bibfield  {author} {\bibinfo {author} {\bibfnamefont {D.}~\bibnamefont
  {Daems}}, \bibinfo {author} {\bibfnamefont {A.}~\bibnamefont {Ruschhaupt}},
  \bibinfo {author} {\bibfnamefont {D.}~\bibnamefont {Sugny}},\ and\ \bibinfo
  {author} {\bibfnamefont {S.}~\bibnamefont {Gu\'erin}},\ }\bibfield  {title}
  {\bibinfo {title} {Robust quantum control by a single-shot shaped pulse},\
  }\href {https://doi.org/10.1103/PhysRevLett.111.050404} {\bibfield  {journal}
  {\bibinfo  {journal} {Phys.~Rev.~Lett.}\ }\textbf {\bibinfo {volume} {111}},\
  \bibinfo {pages} {050404} (\bibinfo {year} {2013})}\BibitemShut {NoStop}%
\bibitem [{\citenamefont {Van-Damme}\ \emph {et~al.}(2017)\citenamefont
  {Van-Damme}, \citenamefont {Schraft}, \citenamefont {Genov}, \citenamefont
  {Sugny}, \citenamefont {Halfmann},\ and\ \citenamefont
  {Gu\'erin}}]{Van-Damme2017}%
  \BibitemOpen
  \bibfield  {author} {\bibinfo {author} {\bibfnamefont {L.}~\bibnamefont
  {Van-Damme}}, \bibinfo {author} {\bibfnamefont {D.}~\bibnamefont {Schraft}},
  \bibinfo {author} {\bibfnamefont {G.~T.}\ \bibnamefont {Genov}}, \bibinfo
  {author} {\bibfnamefont {D.}~\bibnamefont {Sugny}}, \bibinfo {author}
  {\bibfnamefont {T.}~\bibnamefont {Halfmann}},\ and\ \bibinfo {author}
  {\bibfnamefont {S.}~\bibnamefont {Gu\'erin}},\ }\bibfield  {title} {\bibinfo
  {title} {Robust {NOT} gate by single-shot-shaped pulses: {D}emonstration of
  the efficiency of the pulses in rephasing atomic coherences},\ }\href
  {https://doi.org/10.1103/PhysRevA.96.022309} {\bibfield  {journal} {\bibinfo
  {journal} {Phys.~Rev.~A}\ }\textbf {\bibinfo {volume} {96}},\ \bibinfo
  {pages} {022309} (\bibinfo {year} {2017})}\BibitemShut {NoStop}%
\bibitem [{\citenamefont {Zeng}\ \emph {et~al.}(2018)\citenamefont {Zeng},
  \citenamefont {Deng}, \citenamefont {Russo},\ and\ \citenamefont
  {Barnes}}]{Zeng2018}%
  \BibitemOpen
  \bibfield  {author} {\bibinfo {author} {\bibfnamefont {J.}~\bibnamefont
  {Zeng}}, \bibinfo {author} {\bibfnamefont {X.-H.}\ \bibnamefont {Deng}},
  \bibinfo {author} {\bibfnamefont {A.}~\bibnamefont {Russo}},\ and\ \bibinfo
  {author} {\bibfnamefont {E.}~\bibnamefont {Barnes}},\ }\bibfield  {title}
  {\bibinfo {title} {General solution to inhomogeneous dephasing and smooth
  pulse dynamical decoupling},\ }\href
  {https://doi.org/10.1088/1367-2630/aaafe9} {\bibfield  {journal} {\bibinfo
  {journal} {New J.~Phys.}\ }\textbf {\bibinfo {volume} {20}},\ \bibinfo
  {pages} {033011} (\bibinfo {year} {2018})}\BibitemShut {NoStop}%
\bibitem [{\citenamefont {Zeng}\ and\ \citenamefont
  {Barnes}(2018)}]{Zeng2018a}%
  \BibitemOpen
  \bibfield  {author} {\bibinfo {author} {\bibfnamefont {J.}~\bibnamefont
  {Zeng}}\ and\ \bibinfo {author} {\bibfnamefont {E.}~\bibnamefont {Barnes}},\
  }\bibfield  {title} {\bibinfo {title} {Fastest pulses that implement
  dynamically corrected single-qubit phase gates},\ }\bibfield  {journal}
  {\bibinfo  {journal} {Physical Review A}\ }\textbf {\bibinfo {volume} {98}},\
  \href {https://doi.org/10.1103/physreva.98.012301}
  {10.1103/physreva.98.012301} (\bibinfo {year} {2018})\BibitemShut {NoStop}%
\bibitem [{\citenamefont {Güngördü}\ and\ \citenamefont
  {Kestner}(2019)}]{Guengoerdue2019}%
  \BibitemOpen
  \bibfield  {author} {\bibinfo {author} {\bibfnamefont {U.}~\bibnamefont
  {Güngördü}}\ and\ \bibinfo {author} {\bibfnamefont {J.~P.}\ \bibnamefont
  {Kestner}},\ }\bibfield  {title} {\bibinfo {title} {Analytically parametrized
  solutions for robust quantum control using smooth pulses},\ }\bibfield
  {journal} {\bibinfo  {journal} {Physical Review A}\ }\textbf {\bibinfo
  {volume} {100}},\ \href {https://doi.org/10.1103/physreva.100.062310}
  {10.1103/physreva.100.062310} (\bibinfo {year} {2019})\BibitemShut {NoStop}%
\bibitem [{\citenamefont {Dong}\ \emph {et~al.}(2021)\citenamefont {Dong},
  \citenamefont {Zhuang}, \citenamefont {Economou},\ and\ \citenamefont
  {Barnes}}]{Dong2021}%
  \BibitemOpen
  \bibfield  {author} {\bibinfo {author} {\bibfnamefont {W.}~\bibnamefont
  {Dong}}, \bibinfo {author} {\bibfnamefont {F.}~\bibnamefont {Zhuang}},
  \bibinfo {author} {\bibfnamefont {S.~E.}\ \bibnamefont {Economou}},\ and\
  \bibinfo {author} {\bibfnamefont {E.}~\bibnamefont {Barnes}},\ }\bibfield
  {title} {\bibinfo {title} {Doubly geometric quantum control},\ }\bibfield
  {journal} {\bibinfo  {journal} {{PRX} Quantum}\ }\textbf {\bibinfo {volume}
  {2}},\ \href {https://doi.org/10.1103/prxquantum.2.030333}
  {10.1103/prxquantum.2.030333} (\bibinfo {year} {2021})\BibitemShut {NoStop}%
\bibitem [{\citenamefont {Zhu}\ \emph {et~al.}(2020)\citenamefont {Zhu},
  \citenamefont {Chen}, \citenamefont {Jauslin},\ and\ \citenamefont
  {Gu{\'e}rin}}]{Zhu2020}%
  \BibitemOpen
  \bibfield  {author} {\bibinfo {author} {\bibfnamefont {J.-J.}\ \bibnamefont
  {Zhu}}, \bibinfo {author} {\bibfnamefont {X.}~\bibnamefont {Chen}}, \bibinfo
  {author} {\bibfnamefont {H.-R.}\ \bibnamefont {Jauslin}},\ and\ \bibinfo
  {author} {\bibfnamefont {S.}~\bibnamefont {Gu{\'e}rin}},\ }\bibfield  {title}
  {\bibinfo {title} {Robust control of unstable nonlinear quantum systems},\
  }\href {https://doi.org/10.1103/physreva.102.052203} {\bibfield  {journal}
  {\bibinfo  {journal} {Phys.~Rev.~A}\ }\textbf {\bibinfo {volume} {102}},\
  \bibinfo {pages} {052203} (\bibinfo {year} {2020})}\BibitemShut {NoStop}%
\bibitem [{\citenamefont {Dridi}\ \emph {et~al.}(2020)\citenamefont {Dridi},
  \citenamefont {Liu},\ and\ \citenamefont {Gu{\'e}rin}}]{Dridi2020}%
  \BibitemOpen
  \bibfield  {author} {\bibinfo {author} {\bibfnamefont {G.}~\bibnamefont
  {Dridi}}, \bibinfo {author} {\bibfnamefont {K.}~\bibnamefont {Liu}},\ and\
  \bibinfo {author} {\bibfnamefont {S.}~\bibnamefont {Gu{\'e}rin}},\ }\bibfield
   {title} {\bibinfo {title} {Optimal robust quantum control by inverse
  geometric optimization},\ }\href
  {https://doi.org/10.1103/physrevlett.125.250403} {\bibfield  {journal}
  {\bibinfo  {journal} {Phys.~Rev.~Lett.}\ }\textbf {\bibinfo {volume} {125}},\
  \bibinfo {pages} {250403} (\bibinfo {year} {2020})}\BibitemShut {NoStop}%
\bibitem [{\citenamefont {Boscain}\ \emph
  {et~al.}(2002{\natexlab{a}})\citenamefont {Boscain}, \citenamefont {Charlot},
  \citenamefont {Gauthier}, \citenamefont {Gu\'erin},\ and\ \citenamefont
  {Jauslin}}]{Boscain2002}%
  \BibitemOpen
  \bibfield  {author} {\bibinfo {author} {\bibfnamefont {U.}~\bibnamefont
  {Boscain}}, \bibinfo {author} {\bibfnamefont {G.}~\bibnamefont {Charlot}},
  \bibinfo {author} {\bibfnamefont {J.-P.}\ \bibnamefont {Gauthier}}, \bibinfo
  {author} {\bibfnamefont {S.}~\bibnamefont {Gu\'erin}},\ and\ \bibinfo
  {author} {\bibfnamefont {H.-R.}\ \bibnamefont {Jauslin}},\ }\bibfield
  {title} {\bibinfo {title} {Optimal control in laser-induced population
  transfer for two- and three-level quantum systems},\ }\href
  {https://doi.org/10.1063/1.1465516} {\bibfield  {journal} {\bibinfo
  {journal} {J.~Math.~Phys.}\ }\textbf {\bibinfo {volume} {43}},\ \bibinfo
  {pages} {2107} (\bibinfo {year} {2002}{\natexlab{a}})}\BibitemShut {NoStop}%
\bibitem [{\citenamefont {Boscain}\ \emph
  {et~al.}(2002{\natexlab{b}})\citenamefont {Boscain}, \citenamefont
  {Chambrion},\ and\ \citenamefont {Gauthier}}]{Boscain2002a}%
  \BibitemOpen
  \bibfield  {author} {\bibinfo {author} {\bibfnamefont {U.}~\bibnamefont
  {Boscain}}, \bibinfo {author} {\bibfnamefont {T.}~\bibnamefont {Chambrion}},\
  and\ \bibinfo {author} {\bibfnamefont {J.-P.}\ \bibnamefont {Gauthier}},\
  }\bibfield  {title} {\bibinfo {title} {On the {$K+P$} problem for a
  three-level quantum system: optimality implies resonance},\ }\href
  {https://doi.org/10.1023/a:1020767419671} {\bibfield  {journal} {\bibinfo
  {journal} {J.~Dyn.~Control Syst.}\ }\textbf {\bibinfo {volume} {8}},\
  \bibinfo {pages} {547} (\bibinfo {year} {2002}{\natexlab{b}})}\BibitemShut
  {NoStop}%
\bibitem [{\citenamefont {Torosov}\ and\ \citenamefont
  {Vitanov}(2013)}]{Torosov2013}%
  \BibitemOpen
  \bibfield  {author} {\bibinfo {author} {\bibfnamefont {B.~T.}\ \bibnamefont
  {Torosov}}\ and\ \bibinfo {author} {\bibfnamefont {N.~V.}\ \bibnamefont
  {Vitanov}},\ }\bibfield  {title} {\bibinfo {title} {Composite stimulated
  {R}aman adiabatic passage},\ }\href
  {https://doi.org/10.1103/physreva.87.043418} {\bibfield  {journal} {\bibinfo
  {journal} {Phys.~Rev.~A}\ }\textbf {\bibinfo {volume} {87}},\ \bibinfo
  {pages} {043418} (\bibinfo {year} {2013})}\BibitemShut {NoStop}%
\bibitem [{\citenamefont {Bruns}\ \emph {et~al.}(2018)\citenamefont {Bruns},
  \citenamefont {Genov}, \citenamefont {Hain}, \citenamefont {Vitanov},\ and\
  \citenamefont {Halfmann}}]{Bruns2018}%
  \BibitemOpen
  \bibfield  {author} {\bibinfo {author} {\bibfnamefont {A.}~\bibnamefont
  {Bruns}}, \bibinfo {author} {\bibfnamefont {G.~T.}\ \bibnamefont {Genov}},
  \bibinfo {author} {\bibfnamefont {M.}~\bibnamefont {Hain}}, \bibinfo {author}
  {\bibfnamefont {N.~V.}\ \bibnamefont {Vitanov}},\ and\ \bibinfo {author}
  {\bibfnamefont {T.}~\bibnamefont {Halfmann}},\ }\bibfield  {title} {\bibinfo
  {title} {Experimental demonstration of composite stimulated {R}aman adiabatic
  passage},\ }\href {https://doi.org/10.1103/physreva.98.053413} {\bibfield
  {journal} {\bibinfo  {journal} {Phys.~Rev.~A}\ }\textbf {\bibinfo {volume}
  {98}},\ \bibinfo {pages} {053413} (\bibinfo {year} {2018})}\BibitemShut
  {NoStop}%
\bibitem [{\citenamefont {Schraft}\ \emph {et~al.}(2013)\citenamefont
  {Schraft}, \citenamefont {Halfmann}, \citenamefont {Genov},\ and\
  \citenamefont {Vitanov}}]{Schraft2013}%
  \BibitemOpen
  \bibfield  {author} {\bibinfo {author} {\bibfnamefont {D.}~\bibnamefont
  {Schraft}}, \bibinfo {author} {\bibfnamefont {T.}~\bibnamefont {Halfmann}},
  \bibinfo {author} {\bibfnamefont {G.~T.}\ \bibnamefont {Genov}},\ and\
  \bibinfo {author} {\bibfnamefont {N.~V.}\ \bibnamefont {Vitanov}},\
  }\bibfield  {title} {\bibinfo {title} {Experimental demonstration of
  composite adiabatic passage},\ }\href
  {https://doi.org/10.1103/physreva.88.063406} {\bibfield  {journal} {\bibinfo
  {journal} {Phys.~Rev.~A}\ }\textbf {\bibinfo {volume} {88}},\ \bibinfo
  {pages} {063406} (\bibinfo {year} {2013})}\BibitemShut {NoStop}%
\end{thebibliography}

\end{document}